\documentclass[10pt,twocolumn,letterpaper]{article}

\usepackage[pagenumbers]{wacv} %

\usepackage{graphicx}
\usepackage{amsmath}
\usepackage{amssymb}
\usepackage{booktabs}
\usepackage{microtype}

\usepackage{graphicx}
\usepackage{comment}
\usepackage{amsmath,amssymb} %
\usepackage{color}
\usepackage[style=ieee]{biblatex}
\usepackage{multirow}
\usepackage[table, dvipsnames]{xcolor}
\usepackage{hhline,booktabs}
\usepackage[font=small,labelfont=small]{caption}
\usepackage{subcaption}
\usepackage{float}
\usepackage{makecell}
\usepackage{enumitem}
\usepackage{sepfootnotes}

\newcommand{\Lagr}{\mathcal{L}}

\newcolumntype{L}[1]{>{\raggedright\let\newline\\\arraybackslash\hspace{0pt}}m{#1}}

\providecommand{\etal}{\textit{et al}.}
\providecommand{\ie}{\textit{i}.\textit{e}.}

\providecommand{\vs}{\textit{vs}.}
\providecommand{\etc}{\textit{e}.\textit{t}.\textit{c}.}

\usepackage[breaklinks=true,colorlinks,linkcolor=blue,citecolor=blue,bookmarks=false]{hyperref}

\usepackage[capitalize]{cleveref}
\crefname{section}{Sec.}{Secs.}
\Crefname{section}{Section}{Sections}
\Crefname{table}{Table}{Tables}
\crefname{table}{Tab.}{Tabs.}

\begin{document}

\title{Leveraging Bitstream Metadata for Fast, Accurate, Generalized Compressed Video Quality Enhancement}

\author{\normalsize
Max Ehrlich$^{1}$, Jon Barker$^1$, Namitha Padmanabhan$^2$, Larry Davis$^2$, Andrew Tao$^1$, Bryan Catanzaro$^1$, Abhinav Shrivastava$^2$\\
\normalsize
$^1$NVIDA \qquad $^2$University of Maryland, College Park\\
{\tt \footnotesize \{mehrlich, jbarker\}@nvidia.com} {\tt \footnotesize \{namithap, lsdavis\}@umd.edu} {\tt \footnotesize \{atao, bcatanzaro\}@nvidia.com} {\tt \footnotesize abhinav@cs.umd.edu}
}

\maketitle

\begin{abstract}

    Video compression is a central feature of the modern internet powering technologies from social media to video conferencing. While video compression continues to mature, for many compression settings, quality loss is still noticeable. These settings nevertheless have important applications to the efficient transmission of videos over bandwidth constrained or otherwise unstable connections. In this work, we develop a deep learning architecture capable of restoring detail to compressed videos which leverages the underlying structure and motion information embedded in the video bitstream. We show that this improves restoration accuracy compared to prior compression correction methods and is competitive when compared with recent deep-learning-based video compression methods on rate-distortion while achieving higher throughput. Furthermore, we condition our model on quantization data which is readily available in the bitstream. This allows our single model to handle a variety of different compression quality settings which required an ensemble of models in prior work.

\end{abstract}

\vspace{-1.5em}
\section{Introduction}

At its conception, the internet was a medium for the exchange of text data. This has rapidly changed over the last decade to focus on multimedia, and in particular, video~\cite{oentoro_2021,duggan_2020}. Video compression is, therefore, a critical feature of the modern internet. Even short videos are orders of magnitude larger than text data in their uncompressed form and would be impossible to transmit in a timely manner over even a broadband connection.

\begin{figure}[t]
    \centering
    \includegraphics[width=0.47\textwidth]{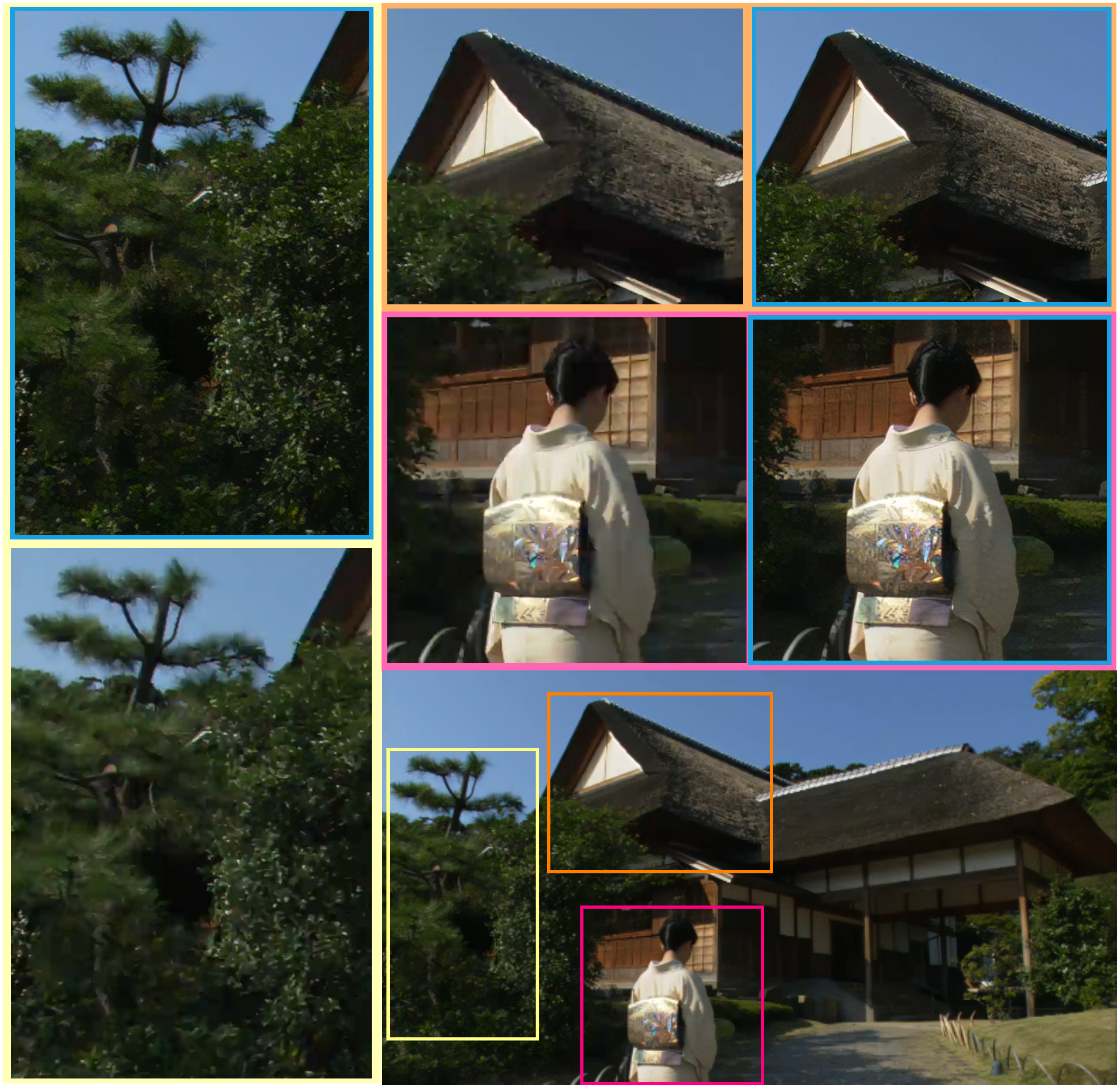}
    \footnotesize
    \caption{\textbf{Don't Spend Megabits, Use MetaBit.} Our MetaBit system takes heavily compressed frames and restores detail. The above example is stored at only 0.039bpp. For each pair, our restoration is shown with a {\color{ProcessBlue}\textbf{blue}} border. Our method is able to faithfully restore natural textures ({\color{Goldenrod}\textbf{left trees}}), clothing textures/human appearance ({\color{RubineRed}\textbf{middle woman}}), and artificial textures ({\color{YellowOrange}\textbf{top roof}}).}
    \label{fig:teaser}
    \vspace{-0.7cm}
\end{figure}

Although modern block-based codecs~\cite{marpe2006h, mukherjee2015technical, sullivan2012overview, chen2018overview, bankoski2011technical} are able to achieve impressive compression ratios with limited quality loss, even these codecs are challenged by bandwidth-limited scenarios which are common in third world countries, rural locations, and lower-class households~\cite{auxier_anderson_2021}. Moreover, the additional transmission latency induced by a low-bandwidth internet connection is unstable: effective bandwidth can vary greatly over time. One way to overcome this limitation is to increase the aggressiveness of the video encoder creating a smaller transmission, however this comes with an associated loss in visual fidelity. We solve this fidelity loss by formulating MetaBit, a novel convolutional neural network architecture \cite{lecun1990handwritten, sutskever2012imagenet} for restoring compressed videos. An example of this is shown in Figure~\ref{fig:teaser}. In effect, we are trading off the unpredictable and often extreme latency of internet transmission for the measurable and predictable latency of deep learning.

Metabit's design is motivated by a number of oversights we identified in prior work. In particular, Metabit is unique in the space of video quality enhancement models because it considers information present in the raw video bitstream. Since the problem we are solving is one caused by video compression, we find it natural to leverage this information which describes in detail how the encoder degraded the given video. We view this metadata as a set of ``hints'' which aid our network in the restoration process. We leverage these hints to improve both the speed and fidelity of the restoration process by directly addressing specific design decisions in prior video quality enhancement models.

\begin{figure}[t]
    \centering
    \includegraphics[width=0.47\textwidth]{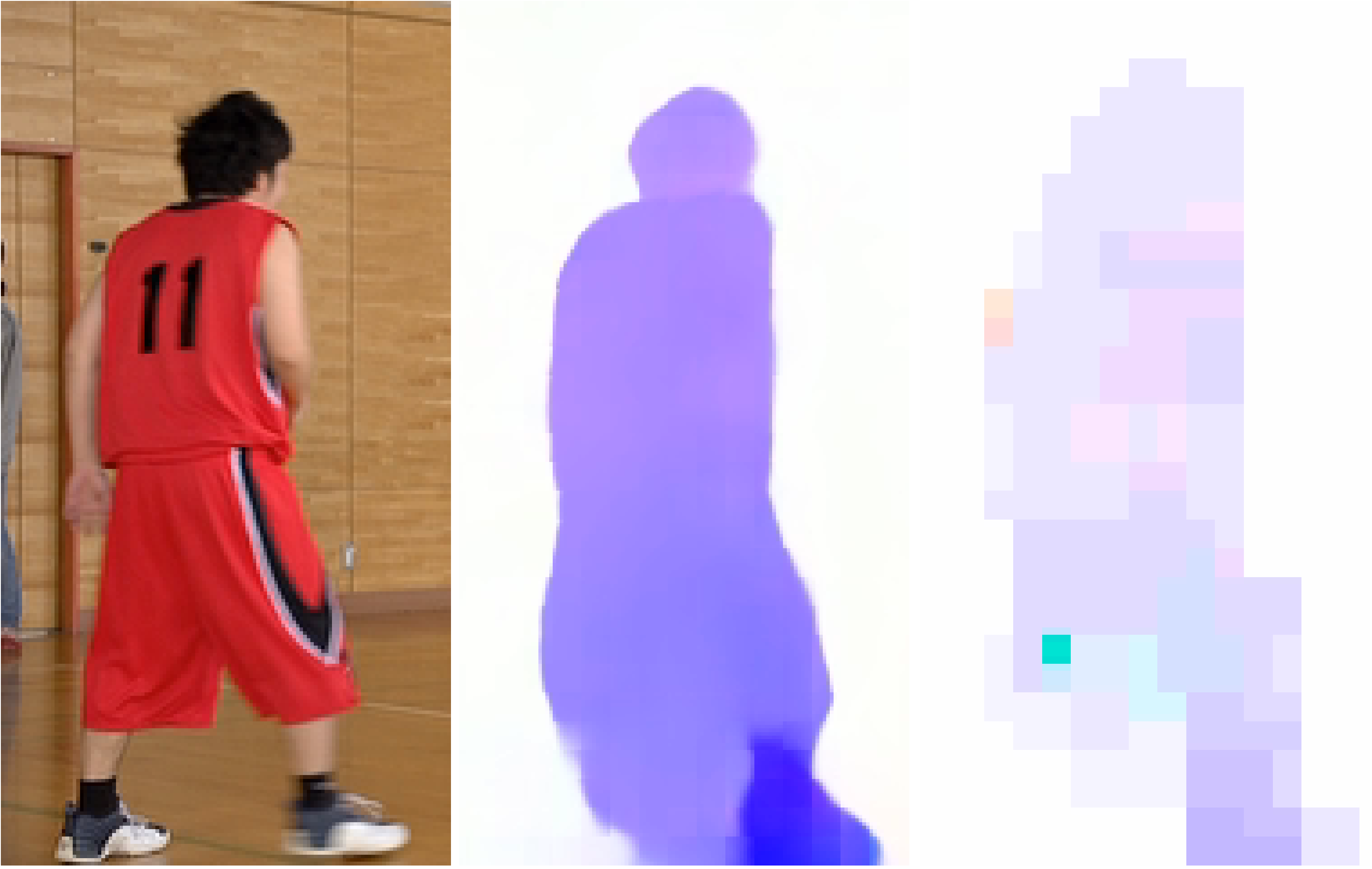}
    \footnotesize
    \caption{\textbf{Motion Vectors.} Motion vectors resemble downsampled optical flow. Left: reference image. Middle: optical flow. Right: motion vectors extracted from the video bitstream. Optical flow was computed with RAFT~\cite{teed2020raft}.}
    \label{fig:mv}
    \vspace{-2em}
\end{figure}

Firstly, prior works expend significant resources on either explicit~\cite{Yang_Xu_Wang_Li_2018,Xing_Guan_Xu_Yang_Liu_Wang_2021} or implicit~\cite{Deng_Wang_Pu_Zhuo_2020, zhao2021recursive, xu2021boosting} motion estimation. This is a resource intensive task for a network which can be entirely avoided by leveraging \textit{motion vectors} from the bitstream that provide the motion information with no computation.  Secondly, although Yang \etal~\cite{Yang_Xu_Wang_Li_2018} correctly observe that not all frames contain the same amount of information, they rely on explicit supervision, and train a discriminative model, to determine the frames with the most information which makes training cumbersome. The idea of using reference frames reappears several times in more recent works \cite{zhao2021recursive, xu2021boosting}. In many cases what these models are using are simply \textit{Intra-Frames} (I-frames) which the encoder intentionally stores at higher quality to use as references for decoding later frames. The position of I-frames is explicitly stored in the compressed bitstream so these high-quality frames can be identified with no computation. Additionally, by differentiating I- and P- frames, we can allocate more parameters to the I-frames and less parameters to the P-frames consequently accelerating the entire architecture. This also introduces a new restoration paradigm: where prior works are sliding-window methods which take a group of frames and produce a single output frame, our method restores blocks of frames at a time, \ie, in a single forward pass, our network consumes 7 degraded frames and predicts 7 high quality frames. This paradigm is much faster than the sliding window methods.

Another critical drawback of all prior works in this space, and one which makes them cumbersome to use in real scenarios, is the requirement that a new model be trained per constant \textit{quantization parameter} (QP) setting. QP settings generally range from 0-51 meaning that in the worst case 51 different models would need to be trained. This also precludes the use of compression methods, like constant bitrate (CBR) and constant rate-factor (CRF), that allow QPs to vary over time and, potentially, space. Since these two compresion methods are in widespread use compared to the constant QP method, prior art is quite difficult to use on real videos. Luckily, the video bistream contains this quantization data as well, so we forumulate a QP cross attention model that reads the potentially time-and-space varying QP map from the bitstream and directs the restoration blocks to adapt their feature maps to varying quantization. This allows a single model, which is easy to train and deploy, to outperform the prior works which depend on an ensemble of models.

Finally, prior works use a limited benchmark of only the H.265 (HEVC)~\cite{sullivan2012overview} compression algorithm with constant QP compression (see Section~\ref{sec:meth:comp} for a detailed discussion). The HEVC codec saw very limited use historically and is all but deprecated by more modern codecs. The legacy H.264 (AVC) codec~\cite{marpe2006h} currently handles approximately 90\% of internet compressed videos~\cite{oentoro_2021}. In addition to being more common, it generates more noticeable degradations especially when paired with CRF quantization (which is the default setting in ffmpeg and is therefore in widespread use). Since the degradations of this codec are more severe, correction models are more useful. Although we benchmark on the HEVC codec for comparison purposes (Appendix A), we additionally report results using the more common AVC codec using CRF encoding (see Section~\ref{sec:res:ecb}) and show that prior works in general fail to generalize to these more complex degradations.

We additionally propose new loss functions. In particular, we formulate a \textit{Scale-Space}~\cite{lowe1999object} loss that allows the network to focus on high frequency details which are removed by compression and we use a GAN~\cite{goodfellow2014generative} loss which enables the network to hallucinate plausible reconstructions. Finally, we extend our comparison to include fully deep-learning-based compression codecs and find that simply using AVC, a widely supported codec, with our restoration network is competitive in terms of rate-distortion and decoding time.

In summary, our contributions are:
\vspace{-0.5em}
\begin{enumerate}
    \item An efficient formulation for video compression correction which leverages the underlying bitstream structure of compressed videos to achieve state-of-the-art performance.
          \vspace{-0.5em}
    \item A method which requires only a single model to handle a range of different quality settings
          \vspace{-0.5em}
    \item A more rigorous evaluation procedure which includes tests on realistic compression settings.
          \vspace{-0.5em}
    \item Improved loss formulations which allow the network to produce plausible reconstructions in extreme compression scenarios.
\end{enumerate}

\begin{figure*}[ht]
    \centering
    \includegraphics[width=\textwidth]{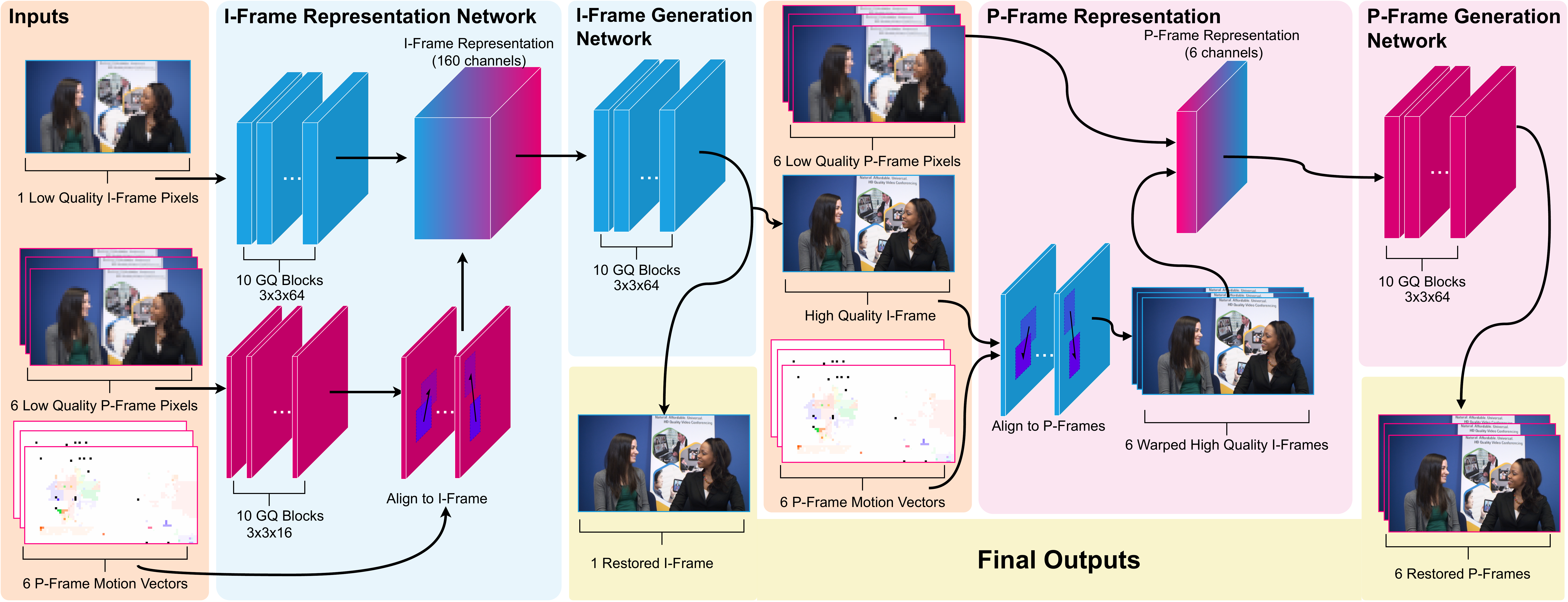}
    \caption{\textbf{MetaBit System Overview.} I-Frames are shown in {\color{ProcessBlue}\textbf{Blue}} and P-Frames are shown in {\color{RubineRed}\textbf{Pink}}. Our network takes an input ({\color{YellowOrange}\textbf{Orange}}) in the form of a low-quality Group-of-Pictures and first performs multi-frame correction on the {\color{ProcessBlue}\textbf{I-Frame}}. The resulting high-quality {\color{ProcessBlue}\textbf{I-Frame}} is used to guide correction of the low-quality {\color{RubineRed}\textbf{P-Frames}}. The final output of our network ({\color{Goldenrod}\textbf{Yellow}}) is the entire high-quality Group-of-Pictures. Please see Section \ref{sec:meth:comp} for an overview of all terminology.}
    \label{fig:overview}
    \vspace{-0.5cm}
\end{figure*}

\vspace{-1em}
\section{Prior Work}

\paragraph{JPEG Artifact Correction}

The related problem of JPEG \cite{wallace1992jpeg} artifact correction is a rich area of study with consistent progress each year. In recent years this problem is solved using convolutional neural networks~\cite{lecun1990handwritten,sutskever2012imagenet}. ARCNN~\cite{dong2015compression} is the first such method which was a simple regression technique inspired by super-resolution architectures. These works were later extended to ``dual-domain'' methods~\cite{liu2015data, zhang2018dmcnn, zheng2019implicit, wang2016d3, liu2018multi}
One flaw in these works was their focus on ``quality-aware'' formulations, in other words. This was solved by Ehrlich \etal~\cite{ehrlich2020quantization, ehrlich2021analyzing} using a formulation which was conditioned on the JPEG quantization matrix and later improved by Jiang \etal~\cite{jiang2021towards} where the network was encouraged to correctly predict the JPEG quality.

\vspace{-0.5cm}
\paragraph{Video Restoration}

Video compression correction is directly related to other video restoration tasks. Toflow~\cite{Xue_Chen_Wu_Wei_Freeman_2019} used optical flow which is trained end-to-end with the restoration task to align frames and operated as a sliding window. EDVR~\cite{Wang_Chan_Yu_Dong_Loy_2019} also operates as a sliding window but replaces the explicit motion estimation with deformable convolutions~\cite{dai2017deformable}. Chan \etal~\cite{chan2021basicvsr} analyze critical components of super-resolution and use this to design a simple, flexible architecture. Relevant here, Li \etal~\cite{Li_Jin_Yang_Liu_Yang_Milanfar_2021} consider super-resolution with common video compression settings. S2SVR~\cite{fgst, lin2022unsupervised} propose a novel unsupervised sequence-to-sequence model which alleviates many of the concerns about sliding-window techniques discussed earlier. Xiao \etal~\cite{xiao2023online} propose kernel grafts as a way to bypass complex networks by transferring their learned representation into a series of lightweight kernels. Lin \etal~\cite{lin2019improved} improve HEVC video decoding by incorporating a superresolution network into the video decoder.
\vspace{-3em}
\paragraph{Video Quality Enhancement}

Video quality enhancement, the task we solve here, was initially solved using ``single-frame'' enhancement methods \cite{wang2017novel, yang2018enhancing} which outperform image-based restoration techniques but use only a single frame at a time. Yang \etal~\cite{Yang_Xu_Wang_Li_2018} propose MFQE which takes multiple frames in a sliding window to correct an entire video sequence. In addition to being the first multi-frame video compression correction model, their key contribution is the concept of \textit{Peak Quality Frames} (PQFs). These are individual frames that have a higher perceptual quality than other nearby frames, and they are identified using a manually trained SVM \cite{cortes1995support}. They combine information from nearby frames using pixel-wise motion estimation and warping. Xing \etal~\cite{Xing_Guan_Xu_Yang_Liu_Wang_2021} extend this idea by replacing the PQF detecting SVM with a BiLSTM \cite{schuster1997bidirectional}. Deng \etal~\cite{Deng_Wang_Pu_Zhuo_2020} use implicit motion compensation with deformable convolutions~\cite{dai2017deformable}. They show that this leads to a more accurate and faster formulation. More recently, Ding \etal~\cite{ding2021patch} design an architecture for capturing adjacent patch information more effectively and Zhao \etal~\cite{zhao2021recursive} use a recurrent hidden state and deformable attention to improve the result. Xu \etal \cite{xu2021boosting} show that STDF performance is improved by intelligently selecting reference frames. In contrast to these techniques, our method requires no motion estimation for alignment and no supervision to determine high-information frames.
\vspace{-1em}

\section{Background}
\label{sec:meth:comp}

Our method leverages concepts from video compression to improve both processing speed and restoration quality over prior works. Note that while we have selected AVC for evaluations, our method depends on information found in all codecs and is equally applicable to VP8/9, AV1, \etc, and nothing presented in this section is codec specific.

\vspace{-1.5em}
\paragraph{Group of Pictures}

Modern video encoders pack information over time into a Group of Pictures (GOP) based on the assumption that over a small time interval motion, and therefore the difference between frames, is small. This yields two types of frames: \textit{Intra-frames} (I-frames), used as reference images, and \textit{Predicted-frames} (P-frames), which require a reference to decode. I-frames are so called because they can be decoded using only information contained within the frame similar to a still-image. P-frames contain two major components: \textit{Motion Vectors} (discussed in the next section) and \textit{Error Residuals}. Error residuals are the difference image between the motion-compensated frame and the true frame. They encode all new information that could not be modeled by motion. As I-frames contain most of the information for a GOP, we allocate more parameters to the representation and generation of the high quality I-frame and use that result to guide restoration of the low-information P-frames.

\vspace{-1.5em}
\paragraph{Motion Compensation}

Video codecs include coarse heuristic motion estimation in the encoding process. These motion vectors are computed on blocks of pixels. See Figure~\ref{fig:mv} for a visual comparison of motion vectors to optical flow. This operation alone compresses blocks of pixels into 4 tuples of source and destination while also reducing the entropy of the error residual. Our network reads and applies these coarse motions for alignment in lieu of pixelwise flows which would need to be computed.

\vspace{-1.5em}
\paragraph{Tuning Quality \vs Bitrate}

Modern codecs provide several methods for tuning the perceptual quality of a video stream. By removing information (which lowers the perceptual quality), the codec is able to further compress the stream resulting in a smaller file. The most common methods are \textit{Constant Rate Factor} (CRF) and \textit{Constant Bitrate} (CBR) with CRF being the default method in many implementations~\cite{tomar2006converting}. In the CRF paradigm, the user presents the encoder with an integer in $[0, 51]$ with higher numbers indicating lower quality. The CRF number is considered a ``proxy'' for perceptual quality. In CBR mode, the encoder is asked to target a specific bitrate in bits-per-second (BPS). This mode is commonly used if a stream is to be transmitted over a connection of known maximum bandwidth. The uncommon \textit{Constant Quantization Parameter} (CQP) method is the mode tested by prior works. In both of the above cases, the encoder converts the user input (CRF or target bitrate) to a set of ``QPs'' which are used to quantize transform coefficients. These QPs generally vary over space and time. In CQP encoding, the user provides a single QP directly and the encoder uses it for all blocks in all frames without regard for information content. Use of this method is generally discouraged since the encoder is no longer making intelligent decisions about which information to keep or discard. However, this method simplifies machine learning solutions because a single QP incurs a predictable degradation. In contrast, varying QPs over space and time (as in CRF and CBR) incur different degradations even within the same frame. Since CQP encoding is so uncommon and discouraged we believe this benchmark is unrealistic and instead study the default CRF encoding.
\vspace{-1em}

\begin{figure}[t]
    \centering
    \includegraphics[width=0.48\textwidth]{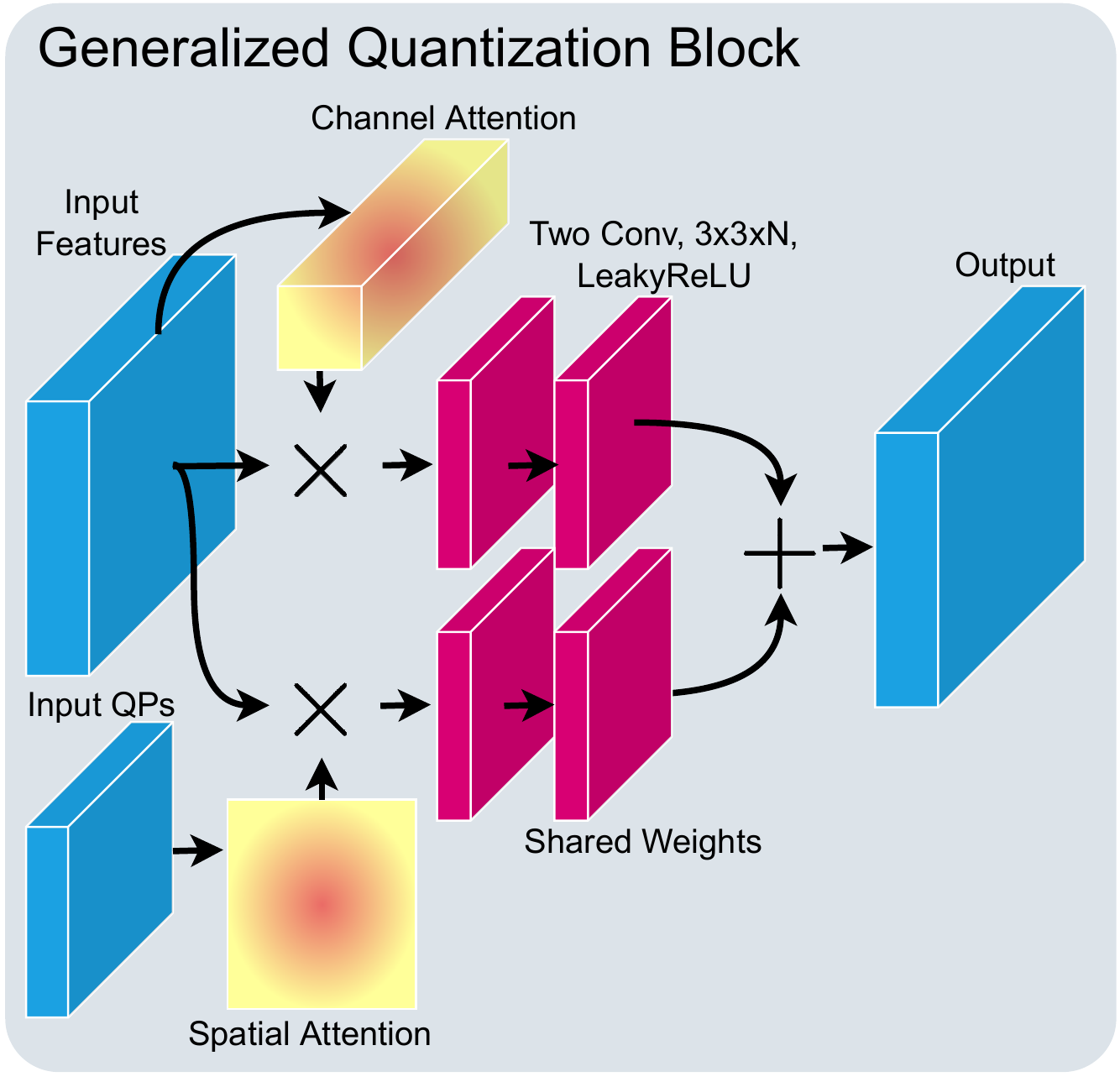}
    \caption{\textbf{Generalized Quantization Block.} Our basic block is designed to be efficient while merging information from the input feature maps and varying quantization data. The block takes the Quantization Parameters (Input QPs) and uses cross-attention to adapt features to varying levels of compression.}
    \label{fig:lrb}
    \vspace{-1.5em}
\end{figure}

\section{Method}

\label{sec:method}

Our task is to take a compressed frame and compute a restored network output which is as close as possible to the target (uncompressed) frame. We accomplish this with a novel multiframe restoration network and loss functions. One unique aspect of compression restoration, when compared to other problems such as denoising, is that we have an exact target frame and we know the exact procedure that was performed on the target frame which caused the degradation. This allows us to design a network architecture that is informed by this procedure, and in our case, incorporate metadata contained in the compressed video bitstream.

In particular, multiframe restoration methods are known to improve when features in different frames are aligned (see Section~\ref{sec:res:abl} for an ablation of this). Rather than compute motion estimation, we leverage the motion vectors contained in the video bitstream. Additionally, video compression allocates more information to I-frames than to P-frames (Section~\ref{sec:meth:comp}). This was reflected in the ``Peak-Quality Frames'' of MFQE~\cite{Yang_Xu_Wang_Li_2018,Xing_Guan_Xu_Yang_Liu_Wang_2021} which required explicit supervision. Instead of computing the locations of these frames, our method simply assumes they are I-Frames\footnote{Examining charts provided by Yang \etal~\cite{Yang_Xu_Wang_Li_2018} shows that the PQFs they detect are likely I-Frames due to their regular spacing, although we do not analyze this here.}. By eliminating the need to perform explicit motion estimation and detect high quality reference frames, we can reinvest the requisite parameters directly in the restoration task leading to high quality results. In the remainder of this section we describe our novel restoration procedure and loss formulations. The procedure is shown graphically in Figure~\ref{fig:overview}.

\subsection{Restoration Procedure}

\paragraph{GQ Blocks}

We build a novel basic-block to perform restoration on video frames with varying quantization. We call this the Generalized-Quantization (GQ) block, which is illustrated in Figure~\ref{fig:lrb}. The design consists of two parallel branches with two convolutional layers each that share weights. The top branch uses channel self-attention to attenuate the most informative input channels. Meanwhile, the bottom branch uses cross-attention computed on the input QP map to adapt the input features to the spatially-varying quantization. The output features are summed to produce the final output of the block. These blocks are stacked to produce the larger structures in our network.

\vspace{-1.5em}
\paragraph{I- and P-Frame Representations}

Our network first performs a multi-frame restoration on I-frames. Since the I-frame itself contains most of the information in a group-of-pictures, we compute a 64-dimensional representation using 10 GQ blocks. We then compute a 16-dimensional representation of the P-frames using 10 GQ blocks. This process is shown in the top part of the blue box in Figure~\ref{fig:overview}.

\vspace{-1em}
\paragraph{Motion Vector Alignment} We then align the P-frame representations to the I-frame representation by warping the P-frame features using motion vectors. Each P-frame contains motion vectors that copy blocks of pixels from the previous frame into new locations in the destination frame. We reverse the direction of these vectors and copy the P-frame features backwards to align with the previous frame. Repeating this process for all motion vectors in the group-of-pictures yields a volume of P-frame features which are coarsely aligned with the I-frame.

\vspace{-1.5em}
\paragraph{I-Frame Generation}

We concatenate the I-frame features and aligned P-frame features channel-wise to yield a volume containing aligned features for the entire group-of-pictures. For a 7 frame GOP, this is a 160-dimensional representation which we project to 64 dimensions for efficiency. We then generate the high-quality I-frame using 10 more GQ blocks with the final one yielding the 3 channel output.

\vspace{-1.5em}
\paragraph{P-Frame Generation}

Finally, we use the high-quality I-frame to generate the high-quality P-frames. Each I-frame is warped using the P-frame motion vectors to yield 6 copies of the I-frame, each one aligned to one of the low-quality P-frames. These warps are concatenated channel-wise with the low-quality P-frames to create a 6 channel input. This is projected to a 64-dimensional feature space and then processed using 10 GQ blocks to yield the high-quality P-frame. This process is shown in the pink boxes of Figure~\ref{fig:overview}.

\begin{table*}[t]
    \centering
    \footnotesize
    \caption{\textbf{Quantitative Evaluation}. We report $\Delta$PSNR (dB) $\uparrow$ / $\Delta$SSIM $\uparrow$ / $\Delta$LPIPS $\downarrow$, averaged over the MFQE \cite{Yang_Xu_Wang_Li_2018} test split.}
    \resizebox{0.7\textwidth}{!}{
        \begin{tabular}{@{}rllll@{}}
    \toprule
    \multicolumn{1}{c}{\multirow{2}{*}{Method}}     & \multicolumn{3}{c}{CRF}                                                                                \\
    \cmidrule{2-4}                                  & 35                               & 40                               & 50                               \\
    \midrule
    MFQE 2.0~\cite{Xing_Guan_Xu_Yang_Liu_Wang_2021} & 0.681 / 0.015 / \phantom{-}0.004 & 0.660 / 0.019 / -0.001           & 0.538 / 0.023 / -0.015           \\
    STDF-R1~\cite{Deng_Wang_Pu_Zhuo_2020}           & 0.862 / 0.011 / \phantom{-}0.032 & 0.814 / 0.015 / \phantom{-}0.030 & 0.632 / 0.023 / \phantom{-}0.013 \\
    STDF-R3L~\cite{Deng_Wang_Pu_Zhuo_2020}          & 0.846 / 0.010 / \phantom{-}0.032 & 0.882 / 0.015 / \phantom{-}0.029 & 0.817 / 0.027 / \phantom{-}0.011 \\
    RFDA~\cite{zhao2021recursive}                   & 0.273 / 0.006 / \phantom{-}0.012 & 0.395 / 0.011 / \phantom{-}0.006 & 0.458 / 0.020 / -0.021           \\
    \textbf{MetaBit (Ours)}                         & \textbf{0.958 / 0.023 / -0.001}  & \textbf{1.032 / 0.031 / -0.018 } & \textbf{0.877 / 0.042 / -0.041 } \\
    \bottomrule
\end{tabular}
    }
    \label{tab:quant}
    \vspace{-0.5cm}
\end{table*}

\subsection{Loss Functions}
\label{sec:meth:loss}

Restoring videos which were subject to extreme compression is a challenging problem. In general, we found that traditional regression losses alone produce a blurry result. This is directly caused by lossy compression's preference for removing high frequency details which is true for both images and videos. We use two loss functions during training in order to solve this problem.

\vspace{-1em}
\paragraph{Regression Loss}

We use the $l_1$ error as our regression loss. For network output $O$ and target frame $T$ we compute
\vspace{-0.1cm}
\begin{equation}
    \Lagr_1(O, T) = ||T - O||_1
\end{equation}

\vspace{-2em}
\paragraph{Scale-Space Loss}

We use a loss based on the Difference of Gaussians (DoG) scale space \cite{lowe1999object}. The DoG is a fast approximation to the Laplacian of Gaussians and as such functions as a band-pass filter. By isolating these frequency bands and weighting their error equally, the network is encouraged to generate images which match in more than just the low-frequency regions. Formally, given network output $O$ and target frame $T$, we compute 4 scales by downsampling $O$, yielding the scale space
\vspace{-0.1cm}
\begin{equation}
    S = \{ O, O_2, O_4, O_8 \}
\end{equation}
where entry $O_s$ is obtained by downsampling $O$ by the factor $s$ in both the width and height. We then compute the DoG by convolving each entry in $S$ with $5 \times 5$ 2D gaussian kernels of increasing standard deviation $\sigma$:
\vspace{-0.1cm}
\begin{equation}
    G(\sigma)_{i j} = \frac{1}{2\pi\sigma^2}e^{-\frac{i^2 + j^2}{2\sigma^2}}
\end{equation}
for kernel offsets $i, j$. For each scale s we compute four filtered images
\vspace{-0.1cm}
\begin{equation}
    \begin{aligned}
        I_{O,s,1} = G(1.1) * O_s \\
        I_{O,s,2} = G(2.2) * O_s \\
        I_{O,s,3} = G(3.3) * O_s \\
        I_{O,s,4} = G(4.4) * O_s
    \end{aligned}
\end{equation}
where $*$ is the discrete, valid cross-correlation operator. We then compute the difference
\vspace{-0.1cm}
\begin{equation}
    \begin{aligned}
        B_{O,s,1} = I_{O,s,2} - I_{O,s,1} \\
        B_{O,s,2} = I_{O,s,3} - I_{O,s,2} \\
        B_{O,s,3} = I_{O,s,4} - I_{O,s,3} \\
    \end{aligned}
\end{equation}
to yield the per-scale frequency bands. This process is repeated for the target image yielding $B_T$. The final loss is then the sum of absolute error between the frequency bands
\begin{equation}
    \Lagr_\text{DoG}(O,T) = \sum_{s \in \{1,2,4,8\}}\sum_{b=1}^3 ||B_{T,s,b} - B_{O,s,b}||_1
\end{equation}

\begin{table}[t]
    \centering
    \footnotesize
    \caption{\textbf{Throughput.} We measure throughput (FPS) on an \textit{NVIDIA GTX 1080 Ti} GPU. Despite having nearly twice as many parameters our network is faster than or on-par with prior works and still able to run on consumer hardware.}
    \begin{tabular}{@{}rlllll@{}}
    \toprule
    Method                  & 240p & 480p & 720p & 1080p & Parameters (M) \\
    \midrule
    MFQE 2.0                & 25.3 & 8.4  & 3.7  & 1.7   & 0.255          \\
    STDF-R1                 & 38.9 & 9.9  & 4.2  & 1.8   & 0.330          \\
    STDF-R3L                & 23.8 & 5.9  & 2.5  & 1.0   & 1.275          \\
    RFDA                    & 24.0 & 6.0  & 2.6  & 1.0   & 1.250          \\
    \textbf{MetaBit (Ours)} & 26.9 & 5.4  & 2.2  & 1.0   & 2.449          \\
    \bottomrule
\end{tabular}
    \label{tab:efficiency}
    \vspace{-1em}
\end{table}

\vspace{-2em}
\paragraph{GAN and Texture Losses}

We use the Wassertein GAN formulation $\Lagr_{\text{W}}(O,T)$ \cite{arjovsky2017wasserstein} with a critic modeled after DCGAN \cite{radford2015unsupervised}, which we modified using the procedure in Chu \etal~\cite{chu2020learning} to introduce temporal consistency (see Appendix B for more GAN details). We include a texture loss \cite{ehrlich2020quantization} which replaces the traditional ImageNet trained perceptual loss with a VGG \cite{simonyan2014very} network trained on the MINC materials dataset \cite{bell2015material}. Intuitively, if the images are encouraged to produce similar logits from this MINC-trained VGG, then it is likely the two images would be classified as the same material and therefore have similar textures. We compare feature maps from layer 5 convolution 3 of this VGG network. Formally:
\begin{equation}
    \Lagr_\text{texture}(O,T) = ||\text{MINC}_{5,3}(T) - \text{MINC}_{5,3}(O)||_1
\end{equation}
\vspace{-3em}
\paragraph{Composite Loss Functions}
This yields the following two loss functions, a regression loss
\begin{equation}
    \label{eq:regloss}
    \Lagr_R(O, T) = \alpha \Lagr_1(O, T) + \beta \Lagr_{DoG}(O, T)
\end{equation}
which is used for regression-only experiments, and a GAN loss
\begin{equation}
    \label{eq:ganloss}
    \begin{aligned}
        \Lagr_\text{GAN}(O, T) = \alpha \Lagr_1(O, T) + \beta \Lagr_\text{DoG}(O, T) + \\
        \gamma \Lagr_\text{W}(O, T) + \delta \Lagr_\text{texture}(O, T)
    \end{aligned}
\end{equation}
where $\alpha, \beta, \gamma, \delta$ are balancing hyperparameters.

\section{Experiments and Results}

\begin{table}[t]
    \centering
    \footnotesize
    \caption{\textbf{GAN Scores}. The GAN loss allows our network to generate more realistic images than the regression loss alone. Regression loss leads to worse FID scores caused by the smooth, textureless appearance of scene elements. Format is FID $\downarrow$ / LPIPS $\downarrow$}
    \begin{tabular}{@{}rll@{}}
    \toprule
    \multicolumn{1}{c}{\multirow{2}{*}{Method}} & \multicolumn{2}{c}{CRF}                                     \\
    \cmidrule{2-3}                              & 40                      & 50                                \\
    \midrule
    AVC (Degraded Input)                        & 67.07 / 0.289           & 152.19 / 0.511                    \\
    MetaBit (Regression)                        & 80.67 / 0.272           & 154.42 / 0.470                    \\
    \textbf{MetaBit (GAN)}                      & \textbf{37.78 / 0.191}  & \textbf{\phantom{0}95.26 / 0.368} \\
    \bottomrule
    \vspace{-3em}
\end{tabular}
    \label{tab:fid}
\end{table}

\begin{figure*}[t]
    \centering
    \includegraphics[width=\textwidth]{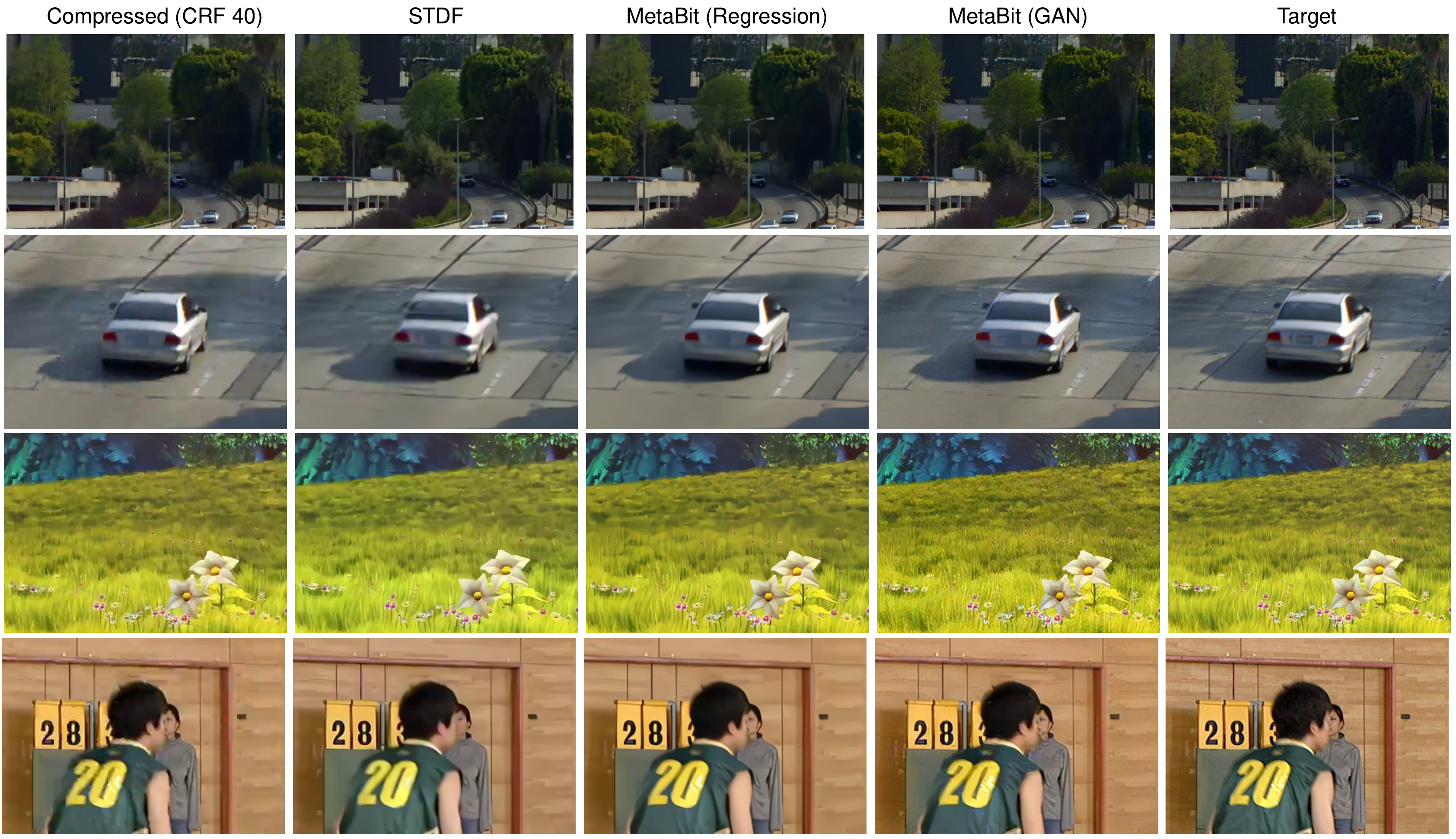}
    \footnotesize
    \caption{\textbf{Qualitative Results.} Please zoom in to view fine details. Note the increased quality of the MetaBit model over STDF and the enhanced sharpness and textures of the GAN method. This is particularly apparent on the trees (row 1), car (row 2), grass (row 3), and the wood texture (row 4). Additional qualitative results are shown in Appendix F and the attached supplement.}
    \label{fig:qual}
    \vspace{-2em}
\end{figure*}

The final architecture as described in Section~\ref{sec:method} contains nearly twice the parameters of the previous state-of-the-art without performing any motion estimation or detection of high quality frames and restores 7 frame blocks. We now show empirically that this formulation works by comparing to prior works on realistic benchmarks. Our method does this while maintaining roughly the same (or better) throughput as the previous methods.

We note two things about our claims here. The first is that the architecture runs similarly-or-faster than models with lower parameter counts because it does not depend on a sliding window and does not need to allocate resources to motion estimation. The second is that this efficient architecture \textit{allows us to allocate more parameters to the model} in order to improve benchmark performance with negligible penalties on speed. This synergizes with our novel loss function and our use of quantization parameters.

\vspace{-1em}

\subsection{Datasets}

We train on the MFQE dataset \cite{Yang_Xu_Wang_Li_2018} training split (108 variable length sequences). We randomly crop $256 \times 256$ patches from each example and apply random horizontal and vertical flipping. We encode the resulting sequence with a 7 frame GOP and no B-frames, thus yielding one I-frame and 6 P-frames per example. We use CRF encoding with auto-variance adaptive quantization for benchmarking. Please see Appendix C for the exact compression commands we used. We evaluate on the MFQE test split. This consists of 18 variable-length sequences (7890 frames) commonly used for evaluation of compression algorithms and was proposed by the Joint Collaborative Team on Video Coding~\cite{ohm2012comparison}.

\subsection{Training Procedure}
\label{sec:res:proc}
Our network is implemented using PyTorch \cite{paszke2019pytorch} and trained end-to-end for 600 epochs using the Adam optimizer \cite{kingma2014adam} with a learning rate of $10^{-4}$. We lower this learning rate to zero over the last 200 epochs using cosine annealing \cite{loshchilov2016sgdr}. For quantitative benchmarks, we train using the regression loss (Equation~\ref{eq:regloss}) with $\alpha=1.0, \beta=1.0$. Note that there is not special balancing that is done in the regression loss formulation.

For GAN training, we begin with regression weights and fine-tune the entire network using our GAN loss (Equation \ref{eq:ganloss}) with $\alpha=0.01. \beta=0.01, \gamma=0.005, \delta=1$. We train for an additional 200 epochs with a learning rate of $10^{-5}$ and the RMSProp optimizer. Please see Appendix B for additional details on the GAN architecture. In this case we use the balancing parameters only to keep the GAN loss from becoming unstable and diverging, this was done by fixing $\delta$, lowering $\alpha, \beta$ together by one order of magnitude, and then lower $\gamma$ until training consistently converged.

For regression we report the change in PSNR, SSIM \cite{wang2004image}, and LPIPS \cite{zhang2018unreasonable} averaged over all frames in the test set. For GAN evaluation, we report the average FID~\cite{heusel2017gans} of the compressed and restored frames.

\begin{figure}[t]
    \centering
    \includegraphics[width=0.43\textwidth]{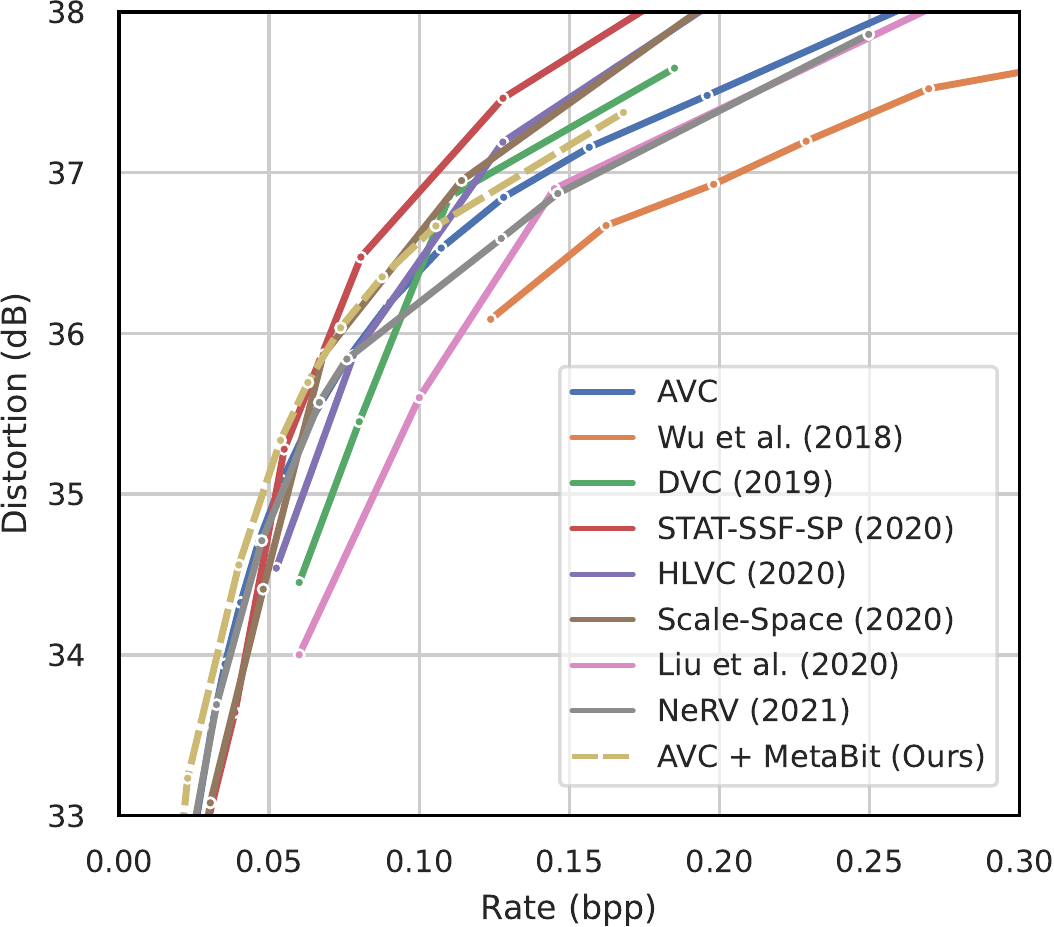}
    \vspace{-1em}
    \caption{\textbf{UVG Rate-Distortion.} Distortion is measured with PSNR. Using AVC + MetaBit method surpasses recent fully deep-learning codecs at low bitrates. As expected, the improvement reduces as bitrate increases. Rate here is ``bits-per-pixel'' (bpp).}
    \label{fig:rd}
    \vspace{-2em}
\end{figure}

\vspace{-1em}
\subsection{Quantitative Results}
We compare to MFQE 2.0~\cite{Xing_Guan_Xu_Yang_Liu_Wang_2021}, STDF~\cite{Deng_Wang_Pu_Zhuo_2020}, and RFDA~\cite{zhao2021recursive}. These methods were all re-trained using publicly available code. We do not compare with single-frame video restoration methods or image restoration methods which were found to have objectively worse performance than the multiframe methods. Please see Appendix A for more details about compared methods and how we chose them. Table~\ref{tab:quant} shows our quantitative results. We test STDF in both the R1 (3-frame sliding window) and the R3 (7-frame sliding window) setting. Of note here is the RFDA result which despite being the most recent compared method made limited gains. This appears to be caused by the larger temporal range that RFDA uses as hidden states are accumulated over time. The temporally varying quality of CRF encoding became a confounding factor.

Overall MetaBit makes an advancement in all metrics over prior works, often by a large margin.  This is particularly noteworthy because we train only a single model to produce all the results in Table~\ref{tab:quant} whereas all prior works required a model per CRF setting. Since our model is conditioned on the QP map, it is able to adapt to spatially and temporally varying degradations in ways that prior works cannot. In particular, MetaBit was the only method which improves perceptual quality (LPIPS) on all three tested CRF values.

We also provide throughput results in Table~\ref{tab:efficiency}. Note that despite having more parameters and better restoration results, our method achieves similar throughput to STDF, even exceeding it in some cases. We are also comparable in speed to MFQE 2.0 which has 9 times fewer parameters. These tests were performed in like-conditions to those reported by prior works to control for the compute environment.

\vspace{-0.8em}
\subsection{GAN Correction}
\label{sec:res:ecb}

While the results on CRF values $\{40, 50\}$ show an improvement in quantitative metrics, these settings represent extreme compression. We found that the regression result of our network is not visually pleasing despite the improvement, so we additionally show results using our GAN procedure. Quantitatively, FID scores in Table~\ref{tab:fid} show a significant improvement in realism with LPIPS scores similarly indicating a significant improvement in perceptual similarity. We show qualitative results in Figure~\ref{fig:qual}, note the significantly improved textures, sharp edges, and additional detail introduced by the GAN loss,  particularly compared to STDF.

\vspace{-0.8em}
\subsection{Comparison to Learned Video Compression}

\begin{table}[t]
    \centering
    \footnotesize
    \caption{\textbf{Compression Throughput.} We measure FPS on an \textit{NVIDIA GTX 2080 Ti} GPU compared to recent deep-learning-based codecs on 1080p frames. For encoding, our method uses AVC and there is no GPU requirement. NeRV \cite{chen2021nerv} encoding speed is not directly reported but requires training a unique network per video.}
    \resizebox{0.48\textwidth}{!}{
        \begin{tabular}{@{}r@{\;}l@{\;}l@{\;}l@{\;}l@{\;}l@{}}
    \toprule
             & Wu \etal \cite{wu2018video} & DVC \cite{lu2019dvc} & Liu \etal \cite{liu2020conditional} & NeRV \cite{chen2021nerv} & \textbf{MetaBit (Ours)}     \\
    \midrule
    Decoding & {\normalsize$10^{-3}$}      & {\normalsize1.8}     & {\normalsize3}                      & \textbf{\normalsize12.5} & \underline{\normalsize3.42} \\
    Encoding & {\normalsize2.4}            & {\normalsize1.5}     & {\normalsize2}                      & {\normalsize-}           & \textbf{\normalsize52}      \\
    \bottomrule
\end{tabular}
    }
    \label{tab:efficiency_lc}
    \vspace{-1.5em}
\end{table}

One application for this work is as a stopgap technology between classical compression and fully deep-learning-based compression. This allows for the speed, memory consumption, and technical debt associated with classical compression algorithms to sustain, with bitstreams fully decodable by users who lack the computational resources for deep models. In Table~\ref{tab:efficiency_lc}, we compare the frame-rate of our method against recently published learned video compression algorithms, and in Figure~\ref{fig:rd}, we compare rate-distortion on the UVG \cite{mercat2020uvg} dataset. Our method is only bested by the recent NeRV~\cite{chen2021nerv} for throughput (which we note has extremely long encoding times) and achieves better rate-distortion results than many compared methods especially at low bitrates.
\vspace{-0.5em}
\subsection{Ablation}

\label{sec:res:abl}

\begin{table}[t]
    \centering
    \footnotesize
    \caption{\textbf{Ablation.} Inference FPS is computed on an \textit{NVIDIA GTX 1080 Ti} for 240p frames, PSNR is computed for H.264 CRF 35.}
    \resizebox{0.5\textwidth}{!}{
        \begin{tabular}{@{}l@{\;}l@{\;}l@{\;}l@{}}
    \toprule
    \multirow{2}{*}{Property}               & \multirow{2}{*}{Option} & \multicolumn{2}{c}{Result}                                                      \\
    \cmidrule{3-4}                          &                         & $\Delta$PSNR (dB)                            & FPS                              \\
    \midrule
    \multirow{2}{*}{Parameter Distribution} & Favors I-Frames         & \textbf{0.954}                               & \textbf{26.9}                    \\
                                            & Even                    & 0.938 ({\color{BrickRed}-1.6\%})             & 24.3 ({\color{BrickRed}-9.7\%})  \\
    \midrule
    \multirow{3}{*}{Motion Compensation}    & Motion Vectors          & 0.948                                        & 26.9                             \\
                                            & Optical Flow            & \textbf{0.952 ({\color{ForestGreen}+0.4\%})} & 17.0 ({\color{BrickRed}-36.8\%}) \\
    \midrule
    \multirow{2}{*}{Loss}                   & $l_1$ and Scale-Space   & \textbf{0.954}                               & -                                \\
                                            & $l_1$ Only              & 0.900 ({\color{BrickRed}-5.6\%})             & -                                \\

    \bottomrule
\end{tabular}
    }
    \label{tab:ablation}
    \vspace{-2em}
\end{table}

We ablate our design in Table~\ref{tab:ablation}, showing impact on throughput and reconstruction accuracy. Note that the first row in each section is the ``reference'' method, \ie, the final model tested in previous sections.

\vspace{-1.5em}
\paragraph{Parameter Distribution} Our architecture allocates more parameters to the I-frame representation than the P-frame representation (64- \vs 16- dimensional). Here, we compare with an ``even'' distribution that allocates a 32-dimensional representation to both. This performs worse in all regards.

\vspace{-0.5cm}
\paragraph{Motion Compensation} We use video motion vectors to perform alignment. A natural comparison is using per-pixel optical flow. For optical flow, we use a pre-trained RAFT~\cite{teed2020raft} model and find that indeed the fine motion detail does improve performance but at a significant throughput penalty.

\vspace{-0.5cm}
\paragraph{Loss} We claimed in Section~\ref{sec:meth:loss} that our scale-space loss helps ensure a correct reconstruction of higher frequency information leading to better reconstruction accuracy. We test this and find that it indeed leads to a 5.6\% improvement over not using a scale-space loss.

\vspace{-1em}

\section{Conclusion and Future Work}

We presented a novel formulation for video compression correction. Our network leverages the structure of the compressed bitstream to outperform prior works while still being extremely efficient. We proposed and tested an improved benchmark with wider applicability. This work has the potential to help people in bandwidth-constrained environments by allowing heavily compressed bitstreams to be viewable.

We hope our work will inspire additional research. Particularly: high-resolution video is slow to process and time-varying compression artifacts can introduce slight temporal inconsistencies (see video examples in the attached supplement). While our parameter count is modest in the space of deep learning models and the method runs on consumer hardware, a practical solution will need to be both smaller and faster. Nevertheless, we believe that this technology is an important stopgap between classical compression and fully deep-learning compression.

\clearpage
{\small
    \printbibliography
}

\clearpage

\appendix

\section{Selection of Compared Methods and HEVC Comparison}

Our paper presents a new and significantly more realistic benchmark than prior works. This required a significant time investment to identify working code releases and retrain compared methods. Some readers may take issue with our particular selection of comparison methods in the body of the paper given that newer methods than those we compared to exist (although we are ``current'' up until last year at the time of writing), so we address this up-front here. When selecting comparison methods, we divided them into three groups:

\begin{description}
    \item[Methods with readily available code] MFQEv2 (2021), STDF (2020)
    \item[Methods with partial code] RFDA (2021)
    \item[Methods with no code] All newer methods (Xu \etal (2021), PSTQE (2021)) and MFQEv1 (2018)
\end{description}

The goal of this exercise is to prioritize which methods we can reasonably compare to. Methods with readily available code can be re-trained and evaluated in a pain-free manner. Methods with partial code required some time investment to correct the code. Methods with no code would require re-implementation which is both time consuming for us and tricky to do in a fair way to the original authors. Optimizing this time-effort-fairness objective precluded us from re-evaluating the methods with no public code.

We note that RFDA has a partial public code release which is not fully functional particularly for retraining (we encourage readers to verify this by visiting the code at \url{https://github.com/zhaominyiz/RFDA-PyTorch}). However, we were able to make the training code functional by filling in details from the paper or correcting the existing code. This was verified by reproducing the paper results to within rounding error reported in the paper. We also note that the PSTQE authors have released the model definition but this alone is not enough to completely train the model even including any training details described in the paper. We opted not to include PSTQE since the training procedure could not be guaranteed to be fair and it would require a significant engineering effort on our part.

In the spirit of further fair comparisons we now show HEVC constant QP results compared to all prior art known to us at the time of writing (we do not claim to have an exhaustive list, nor do we believe a large table necessary to demonstrate that our method works). Since this is the less realistic benchmark which was reported in previous papers we do not have to retrain and we can simply copy the numbers into Table~\ref{tab:pw} allowing for these additional comparisons. To produce our numbers, we followed the same procedure prescribed in prior works by training a separate model per CQP. In this setting, only the recent work by Xu~\etal~\cite{xu2021boosting} is really competitive with our method albeit with the IQE module which adds significant computational overhead. Given that we cannot perform independent testing of their method, it is unclear if that would generalize to more realistic compression benchmarks. It is also unclear to us if our QP cross attention mechanism is contributing in a constructive way to the performance of our model when the QP map is constant for all frames. We respectfully remind the skeptical reader that benchmark results are one aspect of our overall work and are not, were never intended to be, and should not be, a primary contribution.

\begin{table*}[t]
    \centering
    \footnotesize
    \caption{\textbf{HEVC Comparison}. We report $\Delta$PSNR (dB) $\uparrow$ / $\Delta$SSIM $\uparrow$, averaged over the MFQE \cite{Yang_Xu_Wang_Li_2018} test split. This matches exactly with prior works. Note that STDF does not test on QP42. Best in \textbf{bold}, second best \underline{underlined}.}
    \resizebox{0.9\textwidth}{!}{
        \begin{tabular}{@{}rlllll@{}}
    \toprule
    \multicolumn{1}{c}{\multirow{2}{*}{Method}}     & \multicolumn{5}{c}{HEVC CQP}                                                                                                                                            \\
    \cmidrule{2-6}                                  & 22                           & 27                                 & 32                        & 37                                 & 42                                 \\
    \midrule
    MFQE 1.0~\cite{Yang_Xu_Wang_Li_2018}            & 0.31 / 0.0019                & 0.40 / 0.0034                      & 0.43 / 0.0058             & 0.46 / 0.0088                      & 0.44 / 0.0130                      \\
    MFQE 2.0~\cite{Xing_Guan_Xu_Yang_Liu_Wang_2021} & 0.46 / 0.0027                & 0.49 / 0.0042                      & 0.52 / 0.0068             & 0.56 / 0.0109                      & 0.59 / 0.0165                      \\
    STDF-R1~\cite{Deng_Wang_Pu_Zhuo_2020}           & 0.51 / 0.0027                & 0.59 / 0.0047                      & 0.64 / 0.0077             & 0.65 / 0.0118                      & -                                  \\
    STDF-R3~\cite{Deng_Wang_Pu_Zhuo_2020}           & 0.63 / 0.0034                & 0.72 / 0.0057                      & 0.86 / 0.0104             & 0.83 / 0.0151                      & -                                  \\
    RFDA~\cite{zhao2021recursive}                   & 0.76 / 0.0042                & 0.82 / 0.0068                      & 0.87 / 0.0107             & 0.91 / 0.0162                      & 0.82 / 0.0220                      \\
    PSTQE~\cite{ding2021patch}                      & 0.55 / 0.0029                & 0.63 / 0.0052                      & 0.67 / 0.0083             & 0.69 / 0.0125                      & 0.69 / 0.0186                      \\
    Xu~\etal-SQE~\cite{xu2021boosting}              & 0.83 / 0.0046                & 0.92 / 0.0077                      & 0.93 / 0.0116             & 0.85 / 0.0158                      & 0.79 / 0.0218                      \\
    S2SVR~\cite{lin2022unsupervised}                & -                            & -                                  & -                         & 0.93 / 0.0176                      & -                                  \\
    Xu~\etal-IQE~\cite{xu2021boosting}              & \underline{0.96 / 0.0053}    & \textbf{1.09} / \underline{0.0092} & \textbf{1.08 / 0.0136}    & \textbf{1.03} / \underline{0.0190} & \textbf{0.89} / \underline{0.0241} \\
    BasicVSR++ \cite{chan2022basicvsr++}            & 0.90 / 0.0050                & \underline{1.04} / 0.0091          & \underline{1.06} / 0.0128 & \underline{0.99} / 0.0178          & -                                  \\
    \textbf{MetaBit (Ours)}                         & \textbf{1.20 / 0.0109}       & 0.99 / \textbf{0.0117}             & 0.99 / \underline{0.0131} & 0.90 / \textbf{0.0232}             & \underline{0.84} / \textbf{0.0294} \\
    \bottomrule
\end{tabular}
    }
    \label{tab:pw}
    \vspace{-0.5cm}
\end{table*}

\section{GAN Architecture Details}

\begin{figure*}[t]
    \centering
    \includegraphics[width=\textwidth]{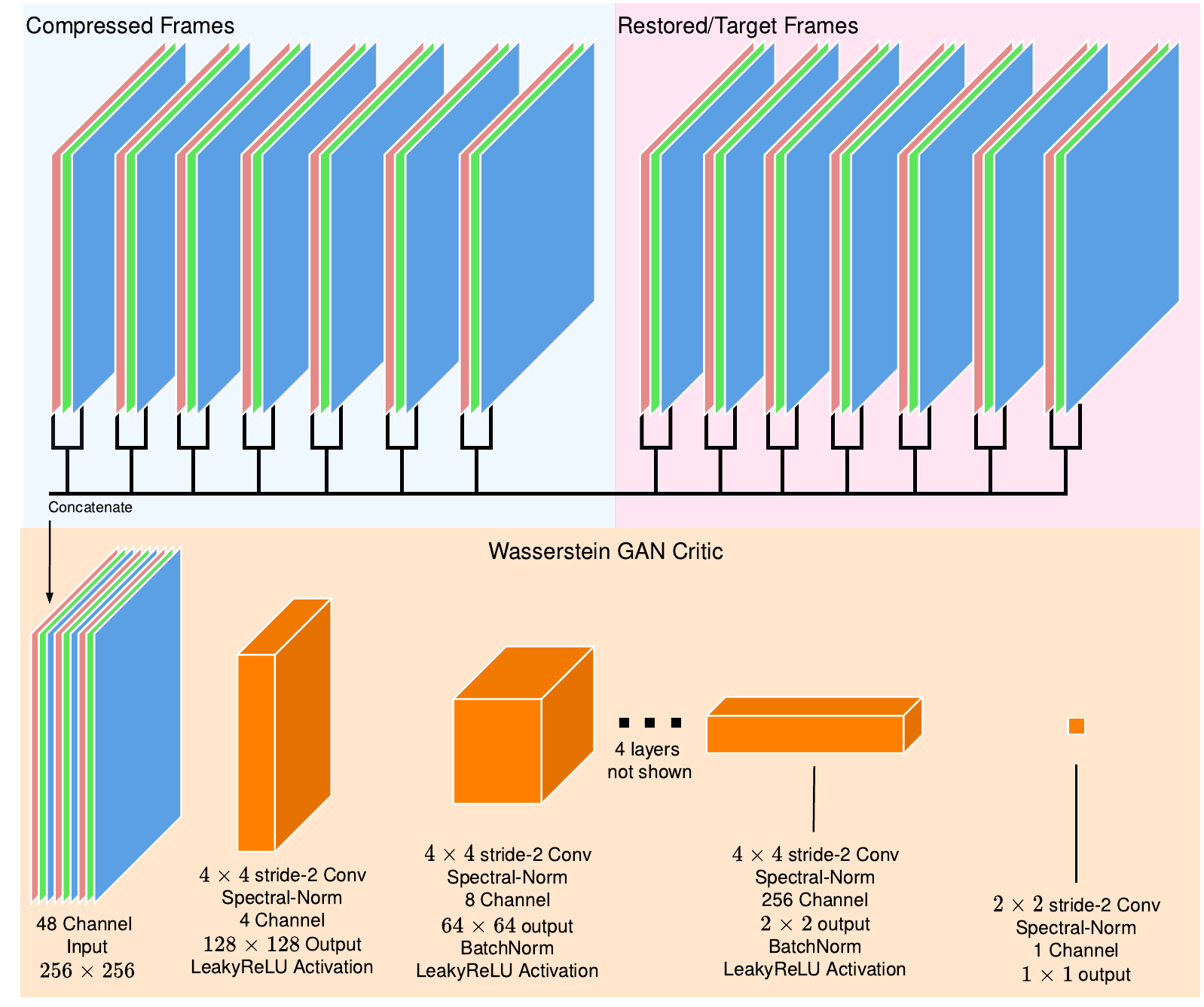}
    \caption{\textbf{Wasserstein GAN Critic.} Inputs (top) are RGB channels for all 7 frames stacked in the channel dimension. The compressed frames are concatenated with the restored or target frames for a 48 channel input. This is then processed by 8 downsampling convolutional layers as in DCGAN to produce the critic output.}
    \label{fig:gc}
\end{figure*}

Chu \etal \cite{chu2020learning} introduce a temporally consistent formulation for video GAN discriminators. We adapt their idea in our GAN loss which is otherwise based on DCGAN \cite{radford2015unsupervised}. The architecture is shown in Figure~\ref{fig:gc}.

Our critic operates on triplets of the compressed frames $C$, network outputs $O$, and target frames $T$. The input frames are stacked channelwise, in other words $C_{0,\text{R}}, C_{0,\text{G}}, C_{0,\text{B}}, ... C_{6,\text{R}}, C_{6,\text{G}}, C_{6,\text{B}}$ for all frames in the GOP are stacked channel-wise along with the corresponding $O$ or $T$ frames to create a 42 channel input (3 channels per frame, times 7 frames, times 2). The task for the critic, then, is to determine whether the $T$ or $O$ channels are network outputs or uncompressed frames given the compressed reference frames and otherwise operates as standard DCGAN. The output is used in a Wassertein GAN loss \cite{arjovsky2017wasserstein}.

This encourages temporal consistency because the critic is judging the entire 7-frame sequence as an example as opposed to other video GANs which treat each frame as an different example. In the later case, there may be situations in which one frame is judged to be more real than another, and yet all frames are equally ``real'' or ``fake''. This ``reallness'' or ``fakeness'' of the sequence \vs the frames is better captured by the formulation of Chu \etal \cite{chu2020learning}. Note that for our task we define ``fake'' as the restored network output and ``real'' as the uncompressed version of the image.

\section{Compression Details}

\begin{figure*}[t]
    \centering
    \includegraphics[width=\textwidth]{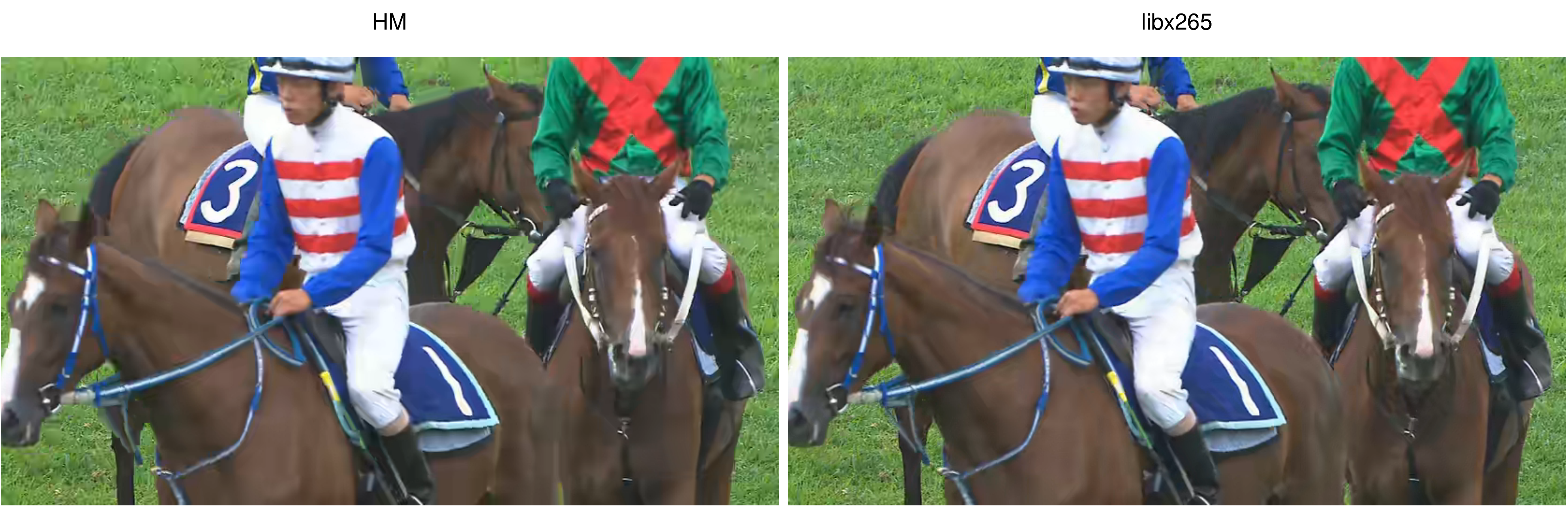}
    \caption{\textbf{Codec Comparison.} HM reference encoder \vs libx265 at QP37. While the artifacts produced by the different encoders are different, the overall perceptual quality is similar. This is likely because of CQP encoding which does not leave many options for the encoder to make intelligent rate-distortion decisions.}
    \label{fig:hmvx265}
\end{figure*}

The MFQE \cite{Yang_Xu_Wang_Li_2018} dataset is stored as a series of uncompressed raw (\texttt{.yuv}) videos of known resolution and frame count. We compress these videos using ffmpeg \cite{tomar2006converting}. For H.264, we use the libx264 encoder with the following command:
\begin{verbatim}
ffmpeg ffmpeg -video_size <WIDTH>x<HEIGHT> \
    -framerate 10 \
    -i <INPUT> \
    -preset medium \ 
    -vcodec libx264 \ 
    -crf <CRF> \ 
    -x264opts <OPTIONS>\
     [OUTPUT]
\end{verbatim}
where \texttt{<OPTIONS>} is defined as:
\begin{verbatim}
keyint=7:min-keyint=7:no-scenecut:no-fast-pskip:me=esa:subme=7:bframes=0:aq-mode=2
\end{verbatim}
This produces a compressed \texttt{.mp4} file for later use (to be read as-is or converted back to raw \texttt{.yuv} for compatibility with prior work). Many of these options are simply there to ensure a 7 frame GOP which is a requirement of the model we presented in the body of the paper (but \textbf{not} a requirement of the general method which we presented). Please note the \texttt{-framerate 10} argument: CRF is sensitive to framerate, so different framerates will incur different degradation strengths. Our choice of 10 was arbitrary and motivated primary by recommendation of Li \etal \cite{Li_Jin_Yang_Liu_Yang_Milanfar_2021}.

For H.264 in ``streaming mode'' (Appendix~\ref{app:sm}) we use the following command in order to be consistent with deep-learning-based compression works:
\\
\begin{verbatim}
ffmpeg -video_size <WIDTH>x<HEIGHT> \
    -framerate 30 \ 
    -i <INPUT> \ 
    -crf <CRF> \
    -preset medium \ 
    -vcodec libx264 \
    -pix_fmt yuv420p \
    -x264opts keyint_min=10000:bframes=0 \
    <OUTPUT>
\end{verbatim}
Here, \texttt{keyint\_min=10000} ensures that there is only a single I-frame per video.

For H.265, prior works evaluated on the HM reference codec, which is notoriously slow. For any models which we retrained on H.265 data, we instead use libx265 which incurs no appreciable change in degradation (Figure~\ref{fig:hmvx265}) and is significantly faster than HM. To generate these videos we used the following command:
\begin{verbatim}
ffmpeg -video_size <WIDTH>x<HEIGHT> \
    -i <INPUT> \
    -vcodec libx265 \
    -qp <QP> \ 
    -x265-params <OPTIONS> \
     <OUTPUT>
\end{verbatim}
for \texttt{<OPTIONS>}:
\begin{verbatim}
keyint=7:min-keyint=7:no-scenecut:me=full:subme=7:bframes=0:qp=<QP>
\end{verbatim}
Note that QP is specified twice and there is no longer a need to control for framerate. We strongly recommend that future works use libx265.

\section{Streaming Mode}
\label{app:sm}

\begin{figure*}[t]
    \centering
    \includegraphics[width=\textwidth]{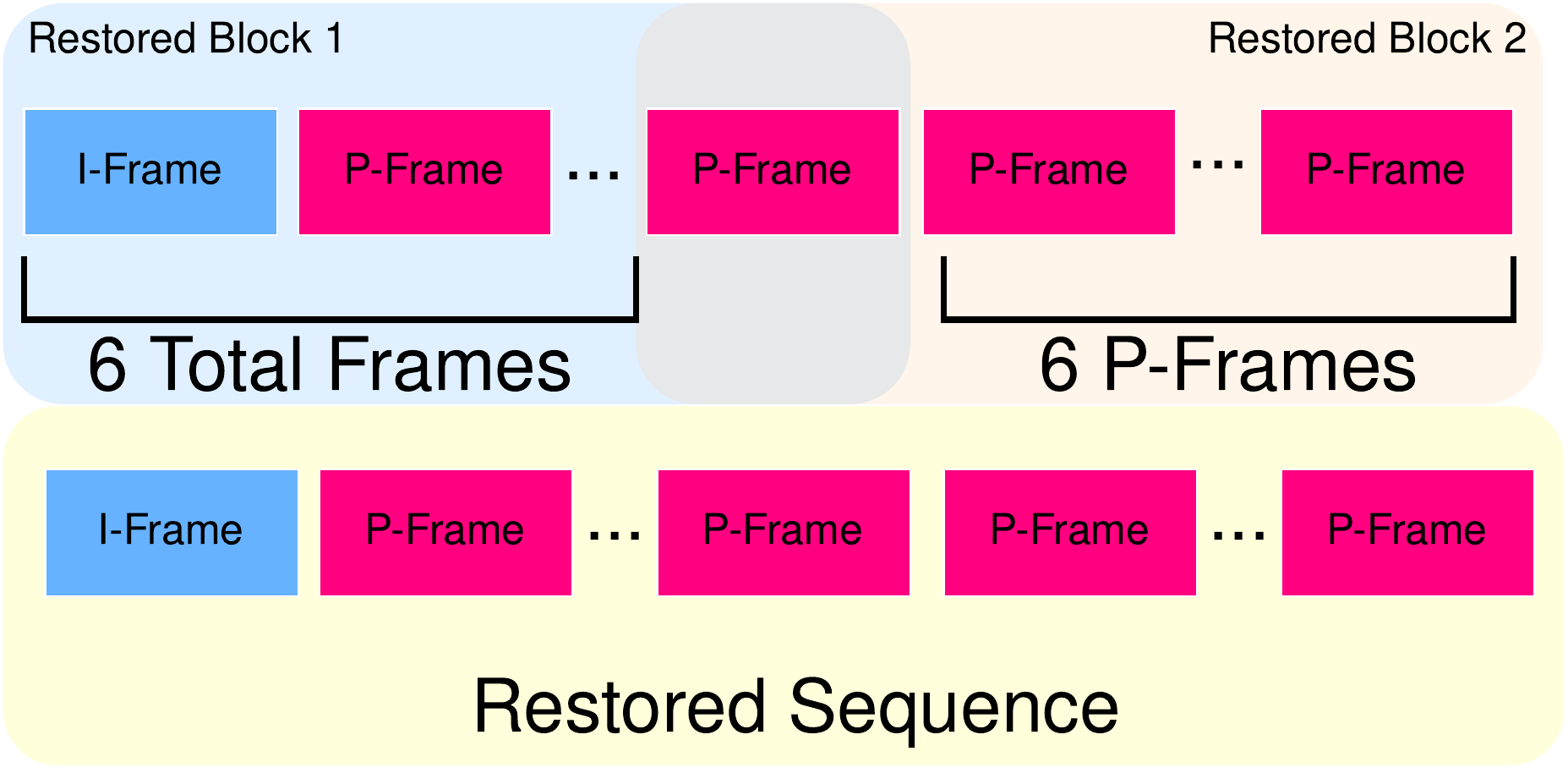}
    \caption{\textbf{Streaming Mode.} In streaming mode, the last restored P-frame in each block is cached and used instead of an I-frame for the next block.}
    \label{fig:sm}
\end{figure*}

In many streaming applications, only a single I-frame is ever transmitted, with every subsequent frame stored as a P-frame. This saves bandwidth and decreases latency at the expense of quality. The model as described in the body of the paper requires a 7-frame GOP with a single I-frame followed by 6 P-frames, however, our model can be easily modified to operate in what we call ``streaming mode''.

In streaming mode (Figure~\ref{fig:sm}), our method performs a 7-frame restoration as usual on the first 7 frames which include the I-frame. The last restored P-frame (frame 7) is cached and used in place of the I-frame for the next 6 P-frames. This process is repeated for all subsequent groups of 6 P-frames, eliminating the need for a periodic high-information frame. This comes at a small cost to restoration performance however it also greatly reduces the bitrate of the H.264 videos. This mode was used to produce the rate-distortion comparisons to deep-learning based compression which requires aggressive settings to match the bitrates reported in the compared works. Adding periodic I-frame incurs a large increase in bitrate. Note that this does not require retraining the model, it is only a change to the inference procedure.

\section{Additional Qualitative Results}

\begin{figure*}[p]
    \centering
    \includegraphics[width=\textwidth]{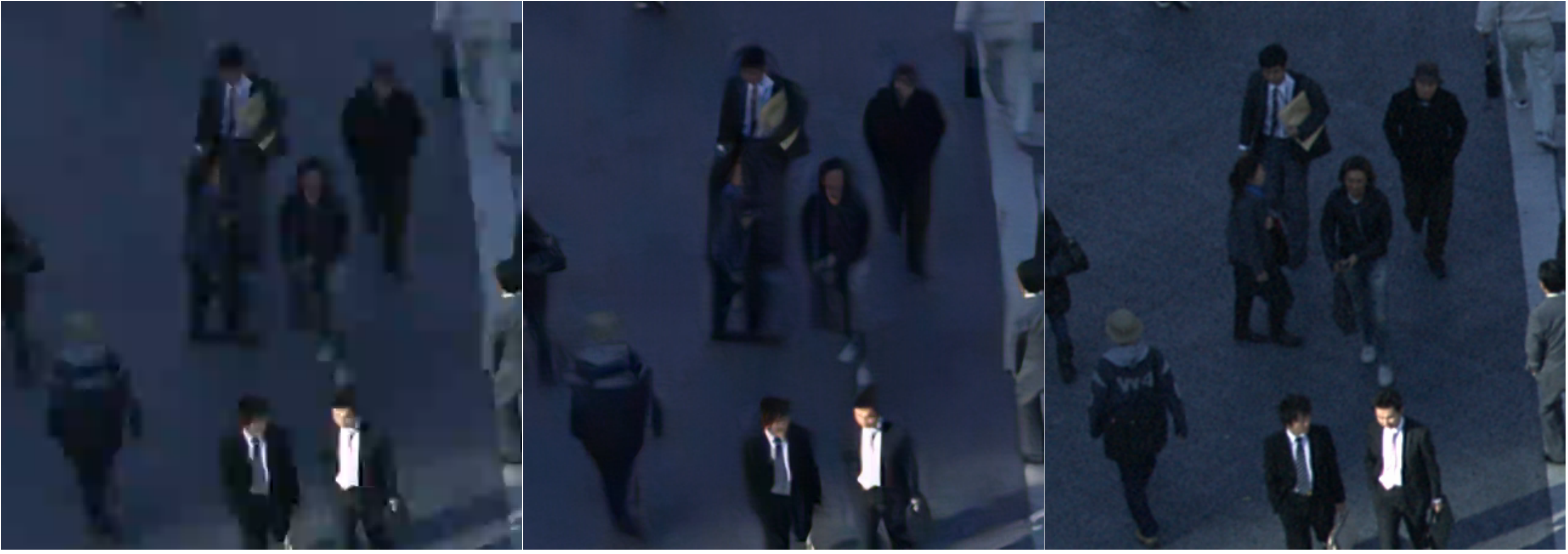}
    \caption{\textbf{Dark Region.} Crop from $2560 \times 1600$ ``People on Street''. The dark region, is poorly preserved by compression. Our GAN restoration struggles to cope with the massive information loss in this region.}
    \label{fig:dark}
\end{figure*}

\begin{figure*}[p]
    \centering
    \includegraphics[width=\textwidth]{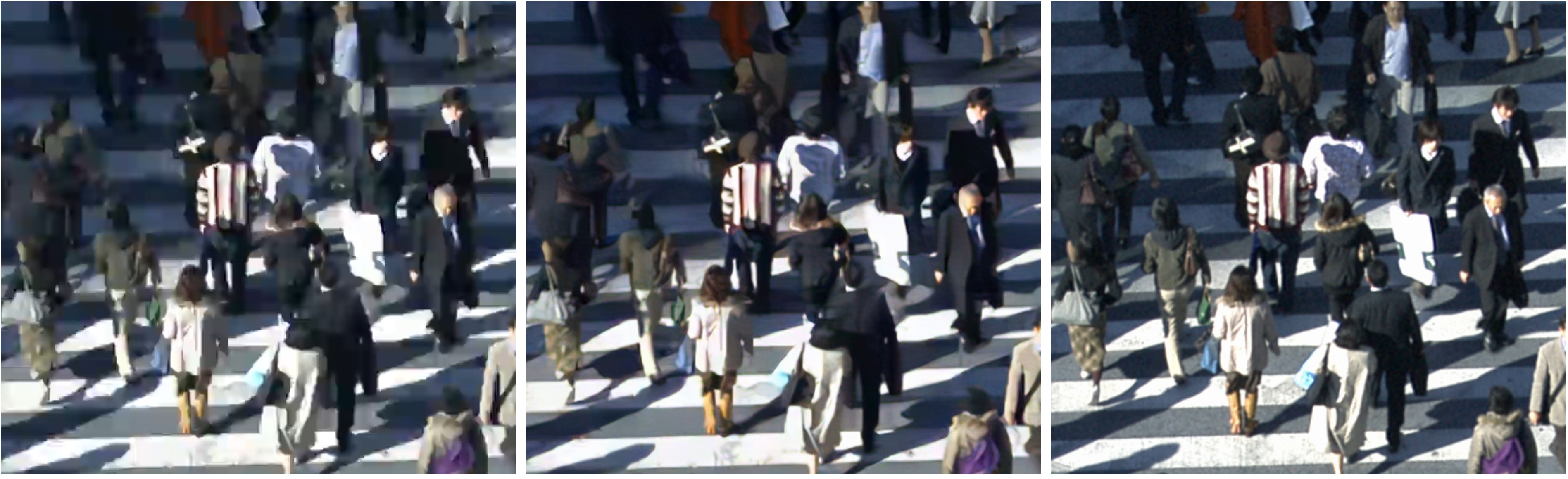}
    \caption{\textbf{Crowd.} Crop from $2560 \times 1600$ ``People on Street''. The image shows an extremely dense crowd. Despite the chaotic nature, our GAN is able to produce a good restoration although there is detail missing.}
    \label{fig:crowd}
\end{figure*}

\begin{figure*}[p]
    \centering
    \includegraphics[width=\textwidth]{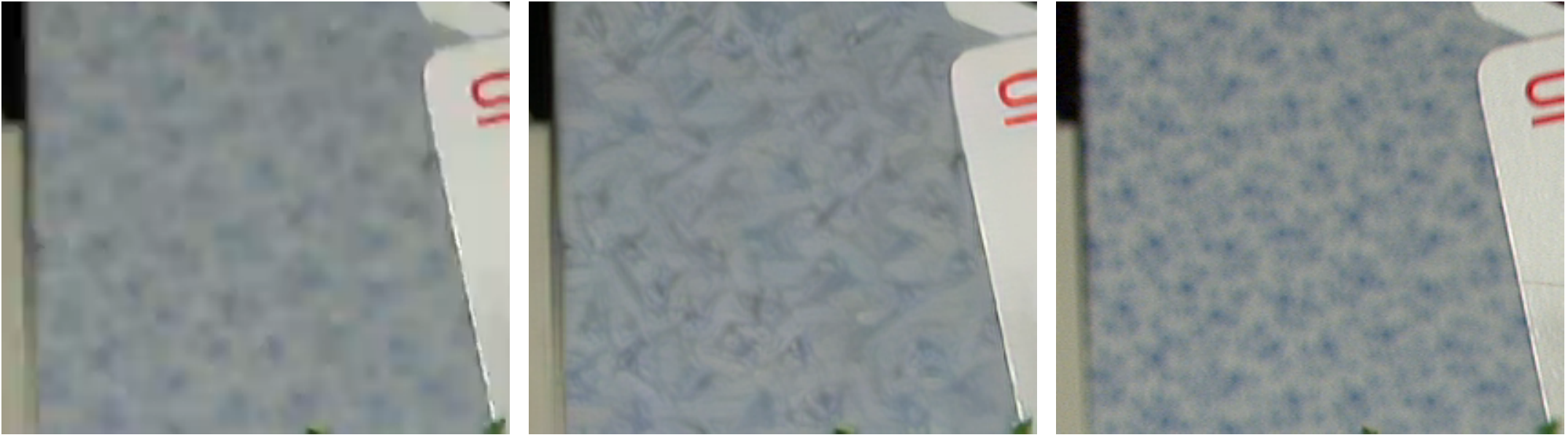}
    \caption{\textbf{Texture Restoration.} Crop from $1920 \times 1080$ ``Cactus''. The texture on the background is destroyed by compression. Our GAN reconstructs a reasonable approximation to the true texture.}
    \label{fig:texture}
\end{figure*}

\begin{figure*}[p]
    \centering
    \includegraphics[width=0.9\textwidth]{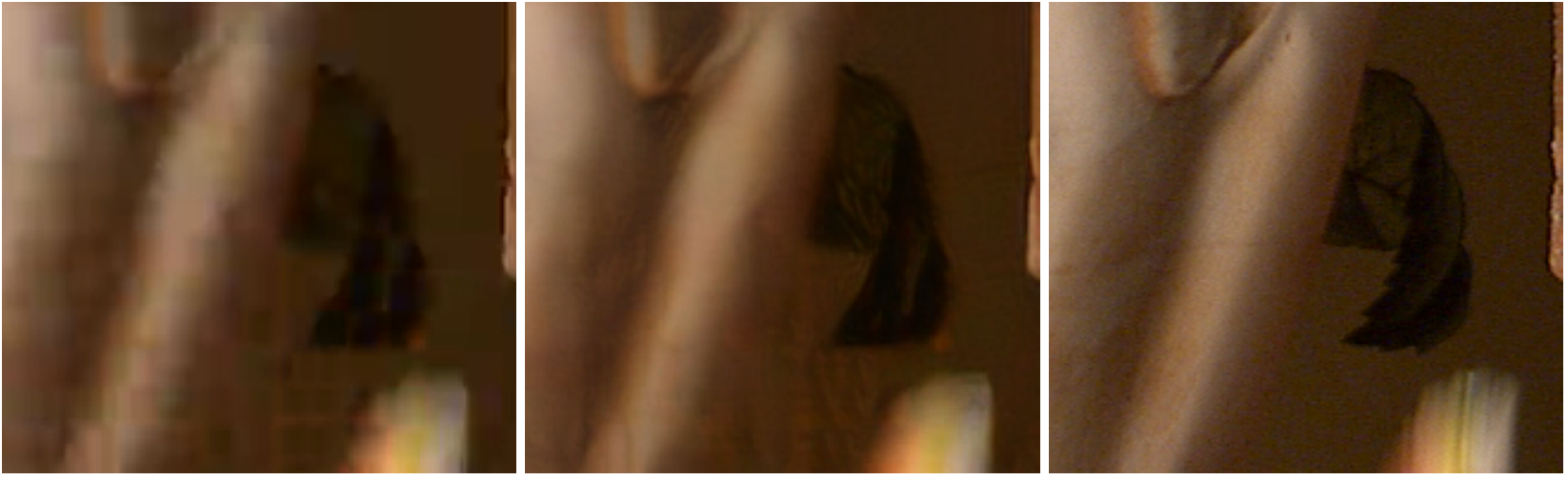}
    \caption{\textbf{Compression Artifacts Mistaken for Texture.} Crop from $1920 \times 1080$ ``Cactus''. The compressed image exhibits strong chroma subsampling artifacts (lower right corner). These are mistaken by the GAN is a texture and restored as such.}
    \label{fig:art}
\end{figure*}

\begin{figure*}[p]
    \centering
    \includegraphics[width=0.9\textwidth]{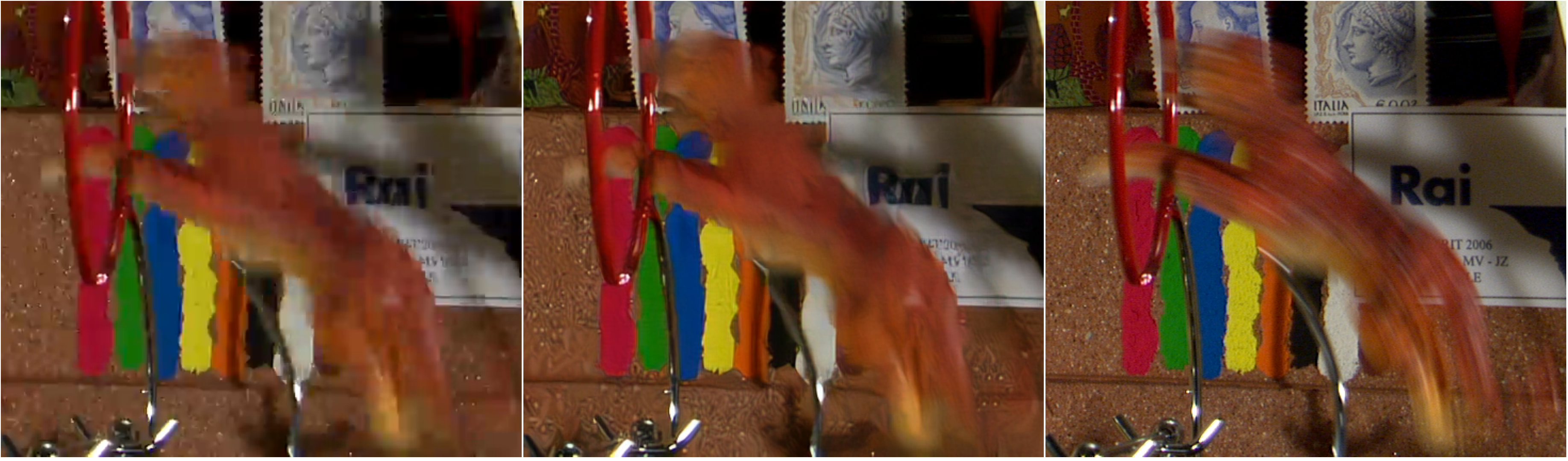}
    \caption{\textbf{Motion Blur.} Crop from $1920 \times 1080$ ``Cactus''. The tiger exhibits high motion which presents itself in the target frame as motion blur. This blur is destroyed by compression and is not able to be restored by the GAN loss. The GAN loss is also ``rewarded'' for sharp edges which would make reconstructing blurry objects difficult. As an aside, note the additional detail on the background objects in the GAN image when compared to the compressed image.}
    \label{fig:motion}
\end{figure*}

\begin{figure*}[p]
    \centering
    \includegraphics[width=0.28\textwidth]{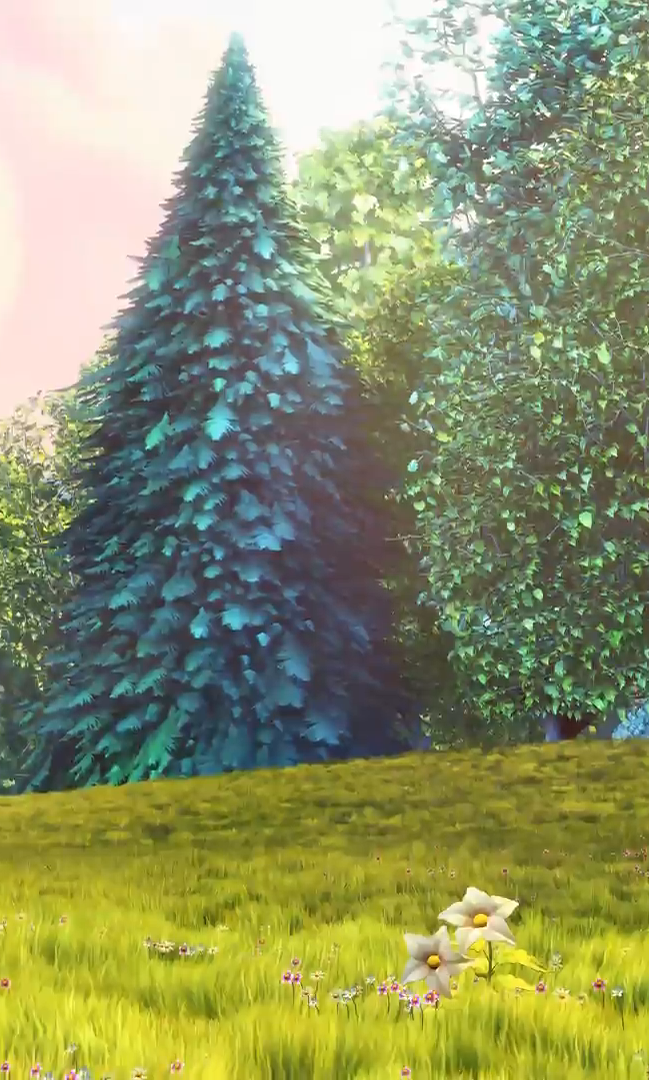}
    \includegraphics[width=0.28\textwidth]{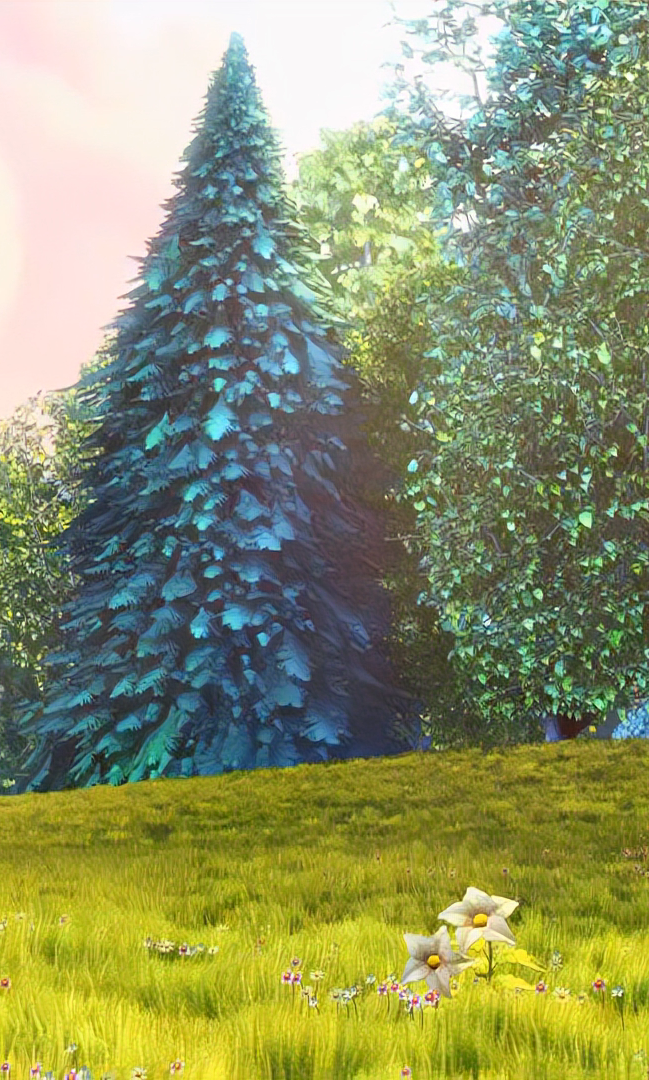}
    \includegraphics[width=0.28\textwidth]{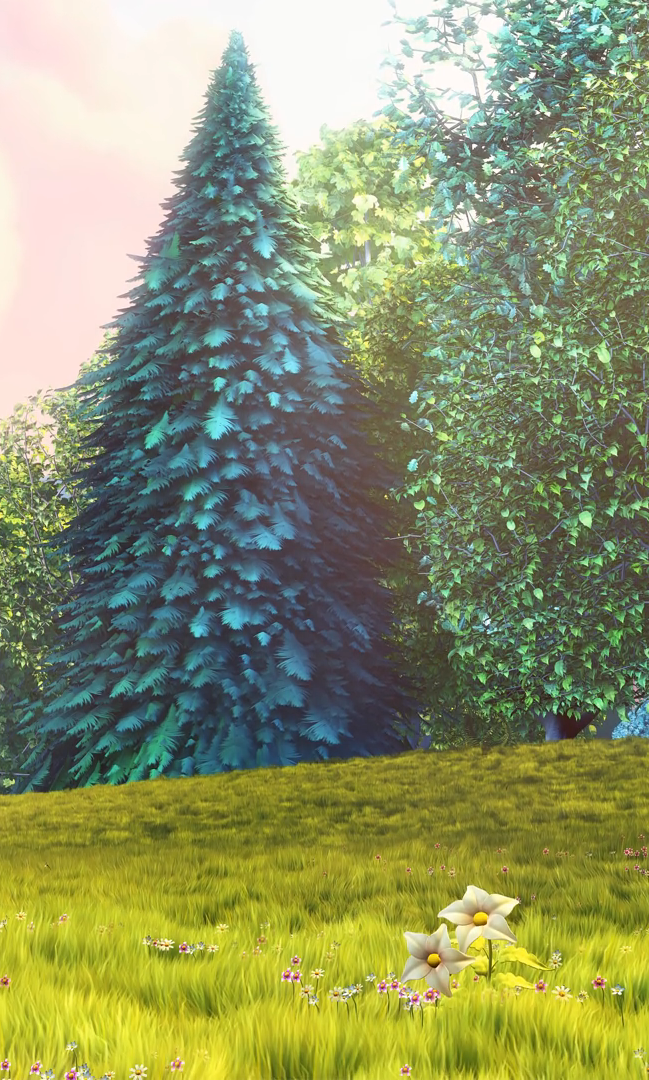}
    \caption{\textbf{Artificial.} Crop from $1920 \times 1080$ ``Big Buck Bunny''. This artificial scene is restored accurately despite a lack of artificial training data. Note the grass and tree textures, sharp edges, removal of blocking on the flower, and preservation of the smooth sky region.}
    \label{fig:bbb}
\end{figure*}

In this section we show additional qualitative results. These results are intended to showcase particular strengths and weaknesses we observed in our model, and are explained further in the figure captions. We encourage viewing the video files contained in our supplementary material to observe temporal consistency issues we noticed due to fluctuating compression artifacts.
\begin{enumerate}
    \item \textbf{Dark Region}~Figure~\ref{fig:dark} highlights a known failure mode of compression causing additional information loss in dark areas in an image.
    \item \textbf{Crowd}~Figure~\ref{fig:crowd} shows our model performance on a dense crowd
    \item \textbf{Texture Restoration}~Figure~\ref{fig:texture} shows an additional result of our model generating a plausible reconstruction of an artificial texture
    \item \textbf{Compression Artifacts}~Figure~\ref{fig:art}; one particular failure mode we observed was compression artifacts, particularly chroma subsampling artifacts, mistaken as a degraded texture and restored by the GAN, creating texture where none exists in the original images
    \item \textbf{Motion Blur}~Figure~\ref{fig:motion}. Another common occurrence is missing motion blur in reconstructed images. There are several issues that lead to this: 1) high motion frames are largely absent from the training data, 2) motion blur is largely destroyed by compression, and 3) the reconstruction loss is explicitly rewarded for generating sharp restorations, whereas in this case we actually \textit{want} a blurry reconstruction.
    \item \textbf{Artificial}~Figure~\ref{fig:bbb}; in this scene from the short film ``Big Buck Bunny'', the frame is restored quite accurately despite a lack of artificial training data.
\end{enumerate}

\end{document}